\documentclass[10pt]{article}   
\pdfoutput=1
\usepackage{jheppub}
\usepackage{amsmath,amssymb,amsfonts,mathbbol,graphicx,slashed,color,amsthm, mathtools, upgreek, enumerate, tensor}

\usepackage{subcaption}
\usepackage{setspace}
\usepackage[export]{adjustbox}
\usepackage{arydshln}
\usepackage[dvipsnames]{xcolor}
\usepackage{physics}
\usepackage{hyperref}
\usepackage{comment}
\usepackage{ulem}
\graphicspath{{pics/}}
\usepackage{breqn}

\usepackage{amsmath}
\usepackage{amssymb}
\usepackage{tikz}

\usepackage{comment}

\usepackage{algorithm}
\usepackage{algpseudocode}

\allowdisplaybreaks

\colorlet{darkblue}{blue!70!black}

\colorlet{darkgreen}{green!50!black}
\colorlet{darkred}{red!50!black}

\newcommand{\mc}{\mathcal}

\def\bea{\begin{eqnarray}}
\def\eea{\end{eqnarray}}
\def\be{\begin{equation}}
\def\ee{\end{equation}}

\title{
 The yes boundaries wavefunctions of the universe
}

\author[a]{Batoul Banihashemi,} \author[a]{Gauri Batra,}
\author[a]{Y.T. Albert Law,}
\author[a]{Eva Silverstein,}
\author[b]{Gonzalo Torroba}
\affiliation[a]{Leinweber Institute for Theoretical Physics at Stanford, 382 Via Pueblo, Stanford, CA 94305}
\affiliation[b]{Centro At\'omico Bariloche, CONICET, and Instituto Balseiro, Bariloche, RN, Argentina}
 
 \vspace{5mm}

\vspace{1cm}

\abstract{
A generic spacetime topology contains timelike boundaries.  Making use of two such boundaries,
we formulate a microscopic holographic dual that captures cosmological spacetime beyond the cosmic horizon patch, including the future wedge.  We build this starting from two copies of the dressed Hamiltonian quantum theories which formulate the cosmic horizon and pole patches of de Sitter.  At the top level of the spectrum we obtain the extended spacetime from a nearly maximally entangled (micro-)canonical thermofield double state.  This requires addressing the maximality of the unrenormalized gravitational path integral saddle in the calculation of the entanglement entropy upon tracing out one sector.  We resolve this in both ensembles via explicit computations in the constrained path integral for three bulk dimensions, incorporating UV-sensitive quantum beyond-GR effects when they contribute strongly.  Lower energy levels in the spectrum generate tall extended spacetimes where the boundaries' causal wedges overlap. These arise in our theory via constraints on the doubled Hilbert space, which encode the operator redundancies arising from the reconstruction of bulk operators from either boundary within the region where their causal wedges overlap.  With positive cosmological constant, the tallness implies that causal wedge reconstruction is more powerful than in the AdS/CFT setting.      
In contrast to the special case of a closed universe, generically quantum gravity with positive cosmological constant---including the future wedge---is manifestly consistent with the existence of multiple states. 
 }

\begin{document}

\maketitle
\parskip=10pt

\section{Introduction}

\begin{figure}[h]
    \centering
    \includegraphics[width=0.7\linewidth]{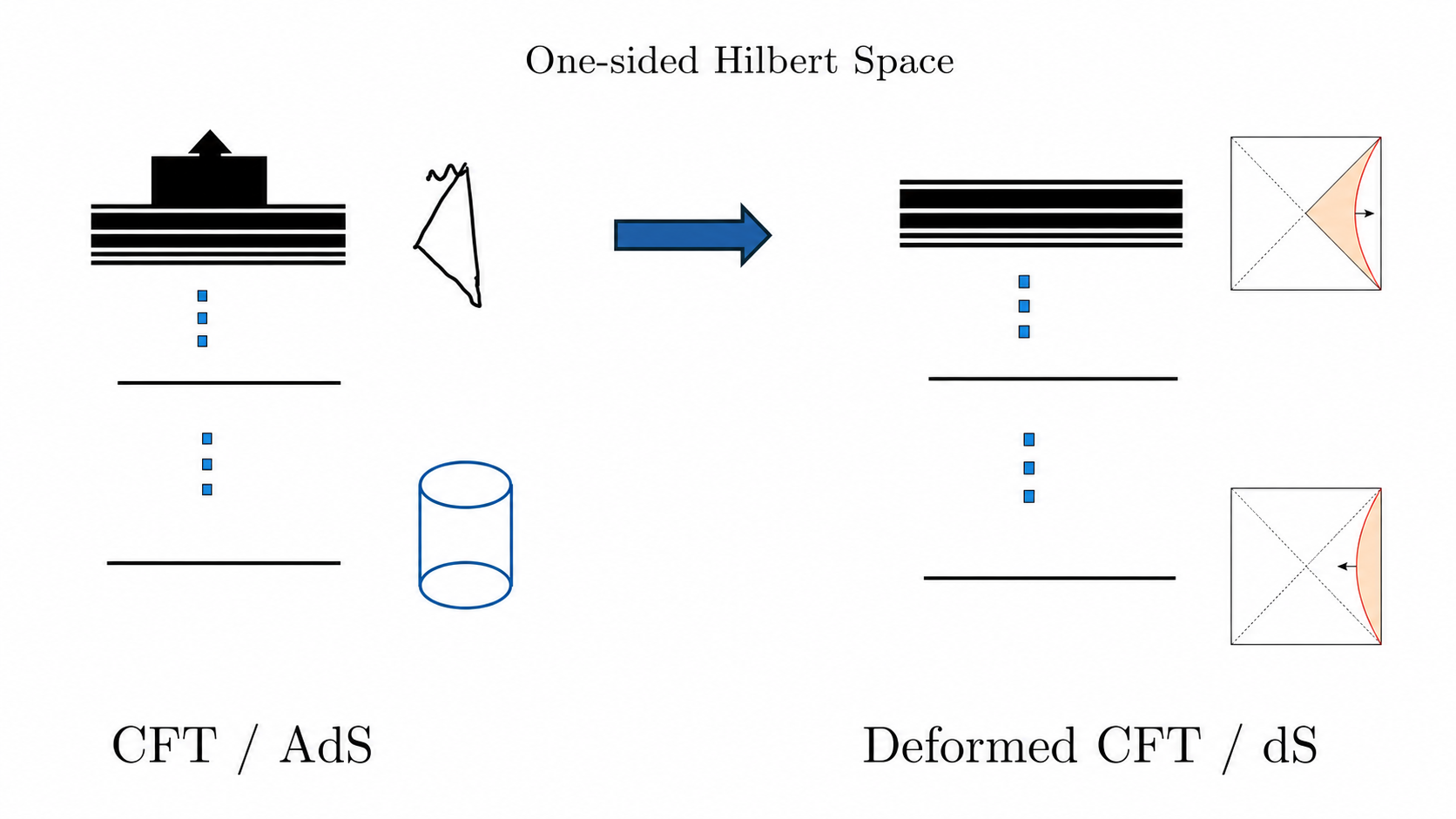}
    \caption{The procedure of \cite{Coleman:2021nor, Batra:2024kjl, Silverstein:2024xnr}  gives a quantum mechanics model whose spectrum includes bands of energy levels dual to the timelike-bounded cosmic horizon patch, pole patch, and local excitations therein (not pictured) which include also landscape decays. This figure depicts the deformation from a CFT to the $T\bar{T}+\Lambda_2$-defomed CFT that is the main aspect of this procedure. The arrows in Penrose diagrams depict the branch choice of the square root in the energy formula (\ref{eq-2d-pure-gr-energy}). This figure is adopted from \cite{Batra:2024kjl}.}
    \label{fig:spectrum}
\end{figure}

Formulating realistic cosmology from first principles is a frontier problem in quantum gravity, with considerable recent progress.\footnote{For a small sample see e.g.~\cite{Anninos:2020hfj, Coleman:2021nor, Chandrasekaran:2022cip, Anninos:2022ujl, Banihashemi:2022htw,Svesko:2022txo, Silverstein:2024xnr,Araujo-Regado:2025elv,Susskind:2021omt,Susskind:2022bia,Susskind:2023rxm,Blacker:2023oan, Susskind:2021esx,Shaghoulian:2021cef, Levine_2023, Tietto:2025oxn,Narovlansky:2023lfz,Narovlansky:2025tpb,A:2023psv, Kawamoto:2023nki,Collier:2025lux,Collier:2024kmo,Aguilar-Gutierrez:2024oea, Aguilar-Gutierrez:2024nst, Maldacena:2024spf,Chen:2025jqm, Chang:2024voo, Abdalla:2025gzn, Chang:2025ays, Anninos:2024wpy, Anninos:2025fer, Anninos:2026hia, Anninos:2024fty,Chakravarty:2024bna, Maxfield:2026ynw, Held:2024rmg, Law:2026tuk, shyam2026dscauchy}, and \cite{Galante:2023uyf, Flauger:2022hie}
for reviews of some aspects.}  
One line of development incorporates timelike boundaries, a generic form of spacetime topology, to formulate a non-gravitational holographic boundary theory \cite{Coleman:2021nor, Batra:2024kjl, Silverstein:2024xnr} {for metastable de Sitter uplifts of AdS/CFT dual pairs \cite{Dong:2010pm, DeLuca:2021pej}}.  With a single timelike boundary, the theories formulate finite spacetime patches such as the timelike-bounded static patch and pole patch along with matter excitations thereof (figure \ref{fig:spectrum}).  The goal of the present work is to build from this to formulate a system capturing larger patches of de Sitter, via a construction with two boundaries (summarized in figure \ref{fig:main_results}). 

In particular,
recent work developed a finite Hamiltonian quantum mechanics system whose Hilbert space includes bands of energy levels dual to bounded patches of semiclassical gravity and matter with positive cosmological constant \cite{Coleman:2021nor, Batra:2024kjl, Silverstein:2024xnr} building from and clarifying \cite{McGough:2016lol, Kraus:2018xrn, Guica:2019nzm, Gorbenko:2018oov, Hartman:2018tkw}.\footnote{Additional holographic models for timelike-bounded systems include the very interesting works \cite{Gross:2019ach, Allameh:2025gsa}.} This gives a microcanonical statistical mechanical realization of bounded de Sitter thermodynamics \cite{Banihashemi:2022htw, Anninos:2020hfj,Anninos:2025zgr,Banihashemi:2025qqi, Draper:2022ofa, Draper:2022xzl,Anninos:2022hqo,Lemos:2024sjs} exhibiting a type I operator algebra at finite $G_N$ (cf \cite{Chandrasekaran:2022cip}).  It explains the meaning of the de Sitter entropy \cite{Gibbons:1977mu} as a count of microstates in a well-defined energy spectrum that exists thanks to the timelike boundary.  It also captures local physics in the bounded cosmic horizon (static) patch of de Sitter spacetime as well as the bounded pole patch to the necessary level of approximation, see figure \ref{fig:spectrum}.  

\begin{figure}[h]
    \centering
    \includegraphics[width=0.5\linewidth]{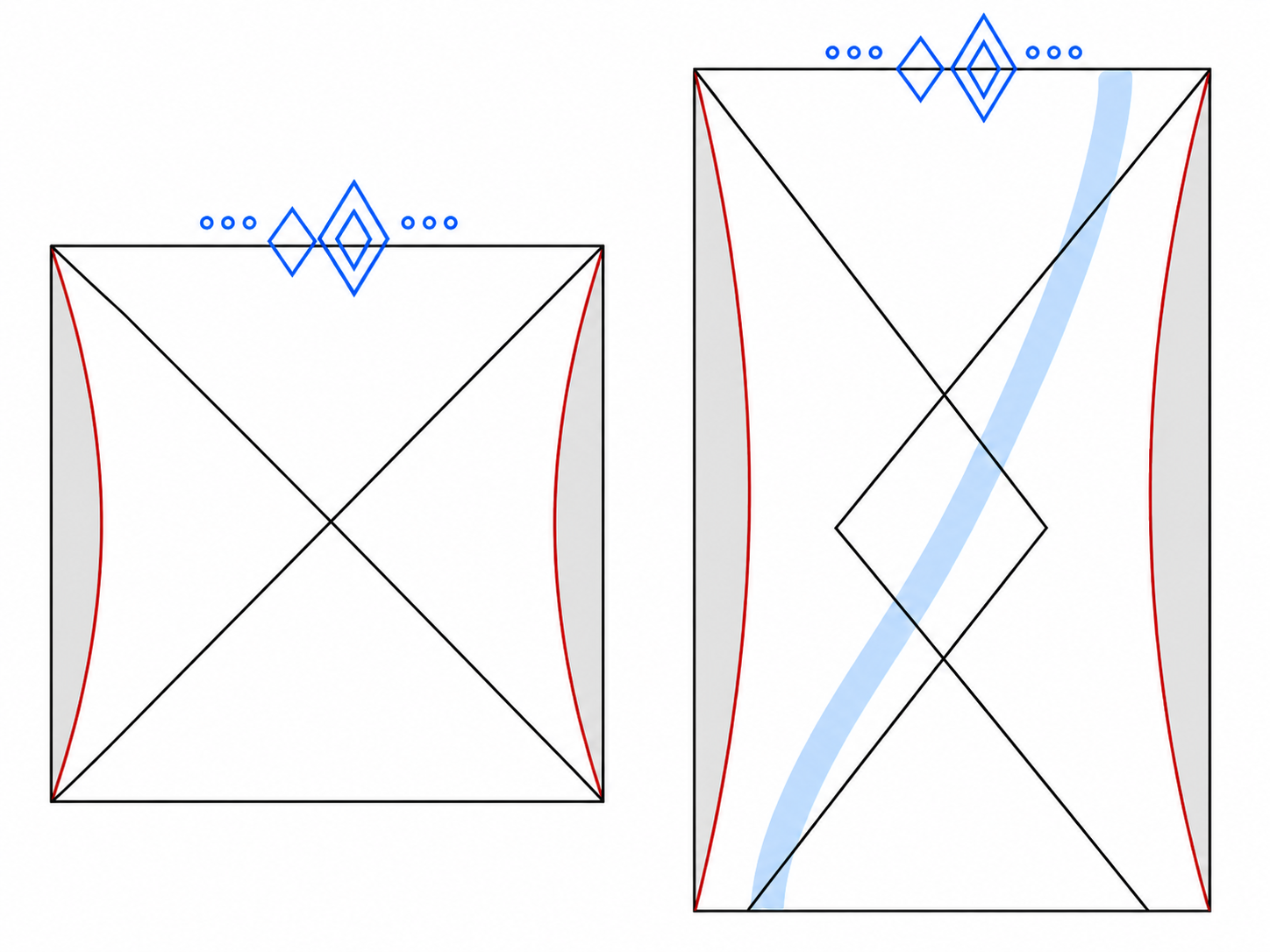}
    \caption{A depiction of two-sided timelike bounded metastable de Sitter space, with a tall geometry with matter on the right. Since the de Sitter vacuum is metastable, the full semiclassical
description also includes Coleman-De Luccia decay histories.  In the
Penrose diagrams here and below we indicate possible decay products schematically by
future diamonds, for example Minkowski diamonds when the terminal vacuum
is asymptotically flat.  These bubbles represent additional histories
whose rates and interiors are determined by the microscopic dS uplift.}
    \label{fig:bounded_dS_tall}
\end{figure}

To capture larger patches of de Sitter, we will focus mainly on a system with two timelike boundaries, whose holographic formulation suffices to capture the extended geometry in the system as depicted in figure \ref{fig:bounded_dS_tall}.  Semiclassical extended states include bare timelike-bounded metastable de Sitter as well as semiclassical states that involve matter.  As we will show, the matter induces joined solutions which are `tall', similarly to the case of global de Sitter \cite{Gao_2000,Leblond:2002ns}.  As a result the two boundaries can communicate.  

One goal of this step in the development of $\Lambda>0$ holography is to understand the role of the de Sitter entropy in the extended system (beyond a single timelike-bounded static patch, where it has already been understood to correspond concretely to a microstate count \cite{Coleman:2021nor, Batra:2024kjl, Silverstein:2024xnr}).  It is important to note that a formulation of the extended geometry including the neck requires less than twice the dS entropy \cite{Giddings:2007nu}.  Although at early and late times effective field theory (EFT) admits many more states than the de Sitter entropy, all but a finite spectrum of states will lead to singularities at finite times.      The consistency of a finite Hilbert space with EFT extends to the Coleman-de Luccia decay processes in the string/M theory landscape as well.    
We will develop a holographic theory that captures the finite spectrum of non-singular geometries, and also includes the individual cosmic horizon microstates in our top band of energy levels (which are generically singular along the horizon).

Another goal in this area is to analyze potential phenomenological implications of non-perturbative cosmological quantum gravity.  There already exists a rich interface between the quantum gravitational structure of string compactification and cosmological observables (see \cite{Silverstein:2016ggb} and references therein), some of which originated from earlier research on holography.  In the present context, one novel outcome has been the realization that topology in the universe may manifest in novel ways \cite{Philcox:2025faf} compared to earlier searches for non-boundary forms of topology \cite{Bond:1999tf,Cornish:2003db,Akrami_2024}.  This suggests analyzing the behavior of the spectrum of ``yes boundaries wavefunctions'' in terms of the probability distribution over the scale of the potential energy generating accelerated expansion, in comparison to the no boundary situation considered in \cite{hartle1983wave,Maldacena:2024uhs, Chen:2020tes, Turiaci:2025xwi,Ivo:2024ill,Abdalla:2026mxn,Cotler:2026wlk,Cotler:2025gui,Nomura:2026igt,Zhao:2026mpl,Antonini:2025ioh, Blommaert:2025bgd,Harlow:2026hky} (see also \cite{Lehners:2023yrj} for a review, or the spacelike boundary situation considered in \cite{Hassfeld:2024hnx}).  More generally, we note a level of apparent optionality in quantum gravity formulations consistent with observation, suggesting a new type of limitation to the phenomenological predictivity of quantum gravity (somewhat analogously to the case of the string landscape).  We defer these directions to future work, motivated by the results in this paper.  

Let us also spell out the terminology used in the title.  By a
``yes-boundary wavefunction'' we mean a gravitational wavefunctional
prepared by a Euclidean path integral with specified timelike boundary
data, in contrast with the usual no-boundary construction in which the
Euclidean geometry closes off smoothly without such boundaries.  From the
field-theory point of view, the  ``yes boundaries wavefunctions''  arise from a UV complete boundary Hamiltonian and a corresponding spectrum of states of
the bounded system. In the
present work the relevant boundary data are the induced metrics, together
with matter boundary conditions, on two timelike walls.  The
resulting wavefunction is therefore a functional of these boundary data
and prepares either the near-maximally entangled top-band state or, for
lower-energy tall geometries, a constrained joined state.  In the pure de
Sitter/top-band limit this construction reduces to the familiar
Hartle-Hawking-type Euclidean preparation, while for tall geometries it
implements the joining constraints associated with the overlapping causal
wedges.

\subsection{Summary of main results and organization of the paper}

The main objective of this work is to generalize the holographic construction of timelike-bounded de Sitter spacetime into non-singular extended spacetime regions beyond the cosmic horizon patch\footnote{Another method is to deform onto a Cauchy slice \cite{Araujo-Regado:2022gvw,  Soni:2024aop, Araujo-Regado:2025elv} in the future wedge \cite{shyam2026dscauchy} somewhat analogously to \cite{AliAhmad:2025kki}.}. 
This includes geometries corresponding both to pure metastable dS bounded by timelike boundaries surrounding the poles, along with tall geometries with matter excitations reviewed with examples in \S\ref{sec-geometries-introduction}.
For the boundary theory, we start with two copies of the Hamiltonian system with finite Hilbert space developed in  \cite{Silverstein:2024xnr, Batra:2024kjl}, and construct a map $\mathcal C$ to the joined system. This proceeds in two steps:

\noindent{{\bf 1.  Top band of energies}}. The joined system contains a `top band' of energy levels near a maximum energy which we analyze in \S\ref{sec-top-joining-TFD-etc}. We construct this band as an approximately factorized product of the corresponding top bands of the one-sided L and R quantum systems. Therefore, for these states the map $\mathcal C$ is approximately the identity. 

At the top of our spectrum, to capture the future wedge of pure metastable dS we ultimately  construct a  microcanonical thermofield double state, which in a constrained path integral framework has no instabilities.  We start, however, by analyzing the canonical thermofield double state at the `tomperature' \cite{Lin:2022nss} of our system  and working through several subtleties it has (cf \cite{Shaghoulian:2021cef}). In the boundary theory, this is analogous to the state that captures two-sided AdS black holes \cite{Maldacena:2001kr}.  

Similarly to that case, starting from our two boundaries the holographic dictionary
instructs us to fill in the bulk geometry and path integrate over its configurations consistent with the boundary conditions. A key difference between the AdS black hole infilling geometry and the dS infilling saddle is that the latter is a maximum of the classical Euclidean action.
Importantly, the path integrand based on the classical gravitational action is not reliable, as we show in several ways. Incorporating necessary UV sensitive effects resolves the tension and provides a derivation of the connectedness of the geometry constructed by the Euclidean path integral.

Given the finite spectrum with the vast majority of states at the top of the spectrum, this state behaves as though it is maximally mixed to good approximation \cite{Dong:2018cuv, Chandrasekaran:2022cip,Lin:2022nss,Banks:2000fe,Bousso:2000nf,Banks:2005bm, Banks:2006rx}.  It admits a microcanonical generalization that focuses on the top band of energy levels.  

The entanglement entropy upon tracing out one of the two identical systems is then explicitly determined to be the de Sitter entropy, a value that matches the Gibbons-Hawking value, i.e.~the area in Planck units (plus interesting corrections analogous to those in \cite{Anninos:2020hfj}).  This matches the dS/dS formulation of the analogous entanglement entropy in \cite{Dong:2018cuv}.\footnote{In that framework, the UV slice playing the role of the boundary of each of the two sectors in \cite{Alishahiha:2004md} automatically lies at the maximal surface, with no room between them to move it off-center in tracing out one of the two systems.}   As just mentioned, we 
address the issue of off-shell contributions which naively provide stronger contributions to the path integral by incorporating UV effects, in the simplest,  $U(1)$-symmetric, setting where the question arises.  We also note more generally that  generalizing Ryu-Takayanagi \cite{Ryu:2006bv, Lewkowycz:2013nqa, Dong:2016fnf, Dong:2023bfy} to off-shell configurations introduces intrinsic UV sensitivity as the conical singularity develops into a curvature singularity \cite{Lewkowycz:2013nqa, Dong:2017xht}.   

\begin{figure}[h]
    \centering \includegraphics[width=0.6\linewidth]{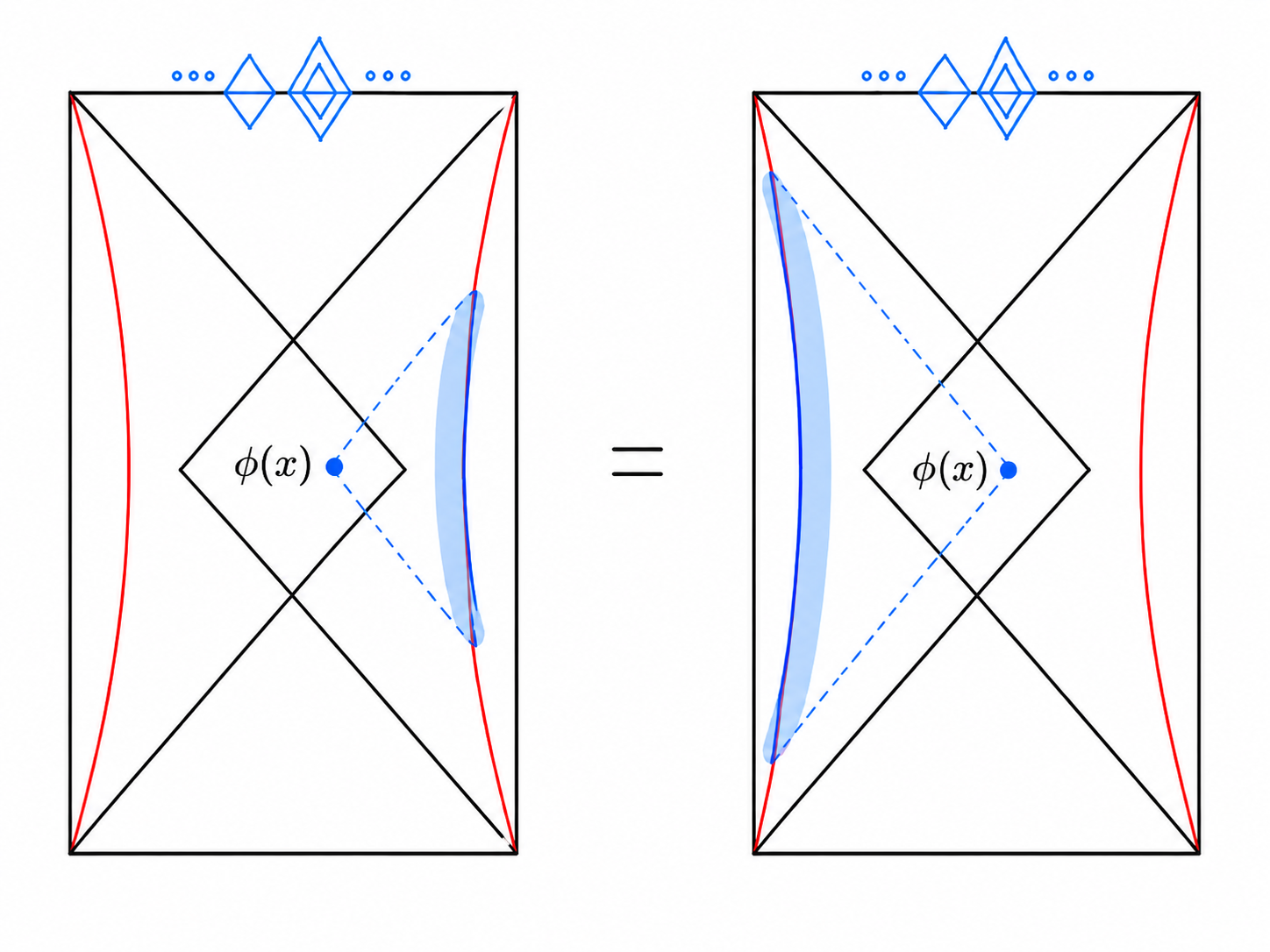}
    \caption{Operators in tall states may be redundantly encoded, either on the shaded region on the left boundary, or on the shaded region on the right boundary. Here the shaded regions in blue depict the regions on the boundary where these operators may be reconstructed. For a more detailed discussion see \S \ref{sec-bulk-reconstruction}.}
    \label{fig:op_redundancy}
\end{figure}

\noindent {{\bf 2. Lower energy tall states}}.  We then determine the structure of the lower-energy states in the joined Hilbert space in \S\ref{sec-constraint-onto-joined-states}.
These states are generically tall similarly to \cite{Gao_2000}.  In order to capture the redundancy of operators in these tall states (see Fig. \ref{fig:op_redundancy}), we introduce constraints ensuring the states are smoothly joined.  At the semiclassical level, the energy spectrum corresponds to the Brown-York quasilocal energy \cite{Brown:1992br}.  The theory we will construct, whose  Hamiltonian can always be written in terms of the energy eigenstates as
\begin{equation}
    H ={\mathcal C} (H_L+H_R){\mathcal C}= \sum E_n | E_n\rangle\langle E_n|
\end{equation}
will by construction match the value of the total Brown-York energy in the semiclassical regime. Gravitationally, tall states can be  constructed in terms of Euclidean ``yes boundary wavefunctions''; these generalize the Hartle-Hawking state (which entangles the top-band one-sided states) to constrained combinations of lighter one-sided states.

Beyond the energy spectrum, the theory captures the dynamics of bulk matter as in \cite{Batra:2024kjl, Silverstein:2024xnr}. In \S \ref{sec-bulk-reconstruction} we study the rudiments of bulk reconstruction via the HKLL \cite{Hamilton:2005ju, Hamilton:2006az} causal wedge reconstruction, formalized algebraically in \cite{Morrison:2014jha, Witten:2023qsv}.  Due to the overlapping causal wedges of the boundaries in tall states, this is more powerful than in the two-sided AdS case, though we also find a new limitation.

\begin{figure}[h]
    \centering \includegraphics[width=1.0\linewidth]{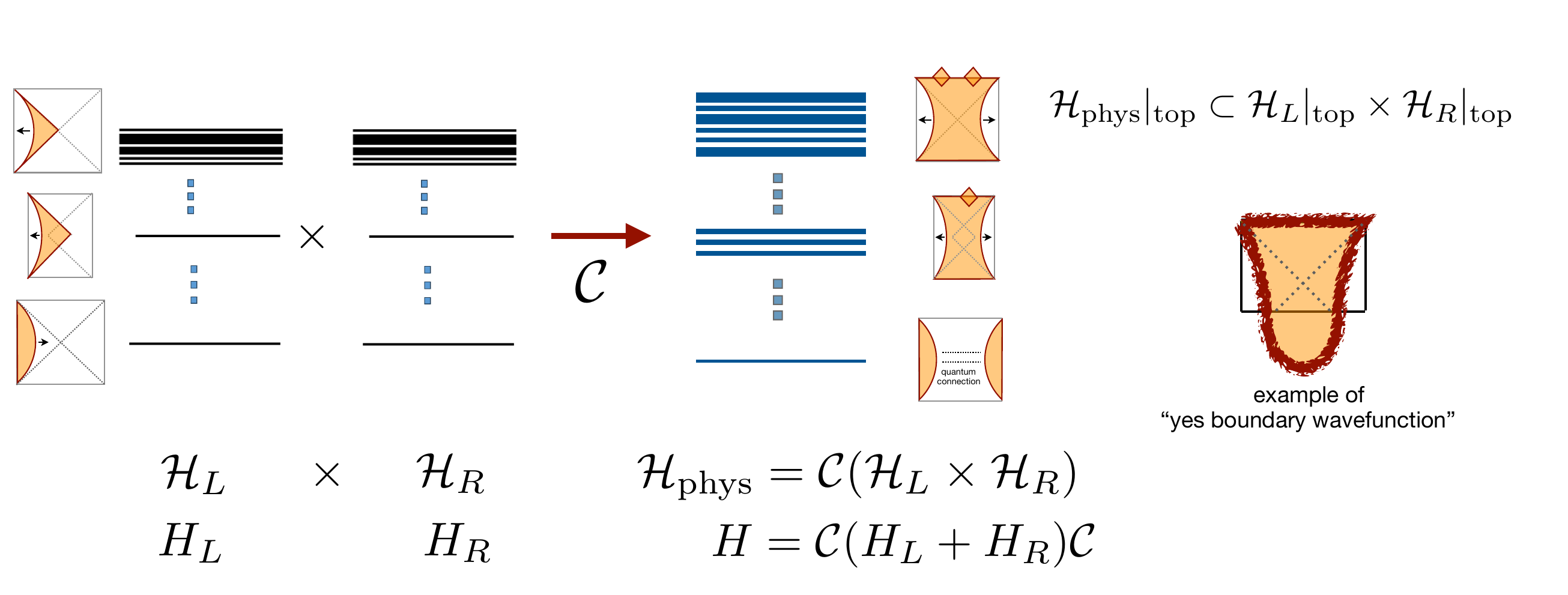}
    \caption[Holographic dual of full de Sitter with Dirichlet walls]{%
    We construct the holographic dual of full de Sitter with Dirichlet walls in terms of a map ${\mathcal C}: {\mathcal H}_L \times {\mathcal H}_R \to {\mathcal H}_{\textrm{phys}}$ from a product of one-sided quantum theories onto a system which includes appropriately joined states even for matter-excited, tall geometries. The physical top band is an approximately factorized space of the corresponding one-sided top bands. The joined geometry depicted in orange arises from nearly maximal entanglement, as we show after resolving the issue of the maximality of the timelike-bounded de Sitter saddle (in configuration space) and the maximality of the horizon (in space). The gravitational results are UV sensitive, and affected by the microscopic nature of the joined spectrum. This is artistically depicted by the fuzziness in the Euclidean wavefunction. Taking this into account we obtain the entanglement and thermal entropy in the joined system. The lower states, including tall geometries, are obtained by imposing constraints that ensure a smooth joining of the L and R states. These constraints can be understood gravitationally in terms of generalized Hartle-Hawking states, comprising additional ``yes boundary wavefunctions.''
    }
    \label{fig:main_results}
\end{figure}

The main features of our construction are summarized in Fig.~\ref{fig:main_results}.

\section{Extended $\Lambda>0$ geometries with timelike boundaries}\label{sec-geometries-introduction}

To set the stage, in this section we introduce relevant timelike-bounded geometries and discuss the semiclassical state space we aim to reproduce holographically.

\subsection{Empty timelike-bounded de Sitter beyond a single cosmic horizon patch}

Among the geometries our theory aims to capture is two-sided timelike-bounded metastable de Sitter
\begin{equation}\label{eq:emptydS}
   ds^2 = -d\tau^2 +\cosh(\tau/\ell)^2 ds^2_{S^3} = -d\tau^2 +\cosh(\tau/\ell)^2 (d\theta^2 + \sin^2(\theta) ds^2_{S^2}) 
\end{equation}
with boundaries at $\theta =\theta_c$ and $\theta = \pi-\theta_c$, with
\begin{equation}\label{eq-boundary-global-coords}
    \sin\theta_c = \frac{R_c}{\cosh(\tau/\ell)}
\end{equation}
along with the decay bubbles as depicted in the figures by Minkowski diamonds at future infinity, see for instance figure \ref{fig:bounded_dS_tall}. 
Here $\ell = \sqrt{3/\Lambda}$ is the de Sitter length and $R_c$ is the fixed boundary proper radius.
Among states that are pure metastable de Sitter in bulk (without matter excitations), the spectrum of states that we will specify also includes the independent microstates of \cite{Coleman:2021nor, Batra:2024kjl, Silverstein:2024xnr}. 
These are generically singular at the horizon.
We will not, however, keep arbitrary singular states, nor will we keep states that correspond to negative mass sources outside the boundaries.

The choice of these two boundaries is part of the definition of the
gravitational and dual quantum mechanical system studied in this paper.  We take them to be spherical
timelike Dirichlet walls of fixed area radius $R_c$; equivalently the
two walls that appear as fixed-radius boundaries in the corresponding
static patches.  In the global coordinates above this fixed-area-radius
condition gives $\sin\theta_c=R_c/a(\tau)$.  The induced boundary
metric, including its lapse or Euclidean periodicity, is the Dirichlet
data specifying the gravitational ensemble.  This is the boundary choice
for which the one-sided finite Hamiltonian systems of
\cite{Coleman:2021nor, Batra:2024kjl, Silverstein:2024xnr} were constructed, and it gives a well-defined Brown-York energy used below. Other choices of timelike boundary data are possible.  They would define
different yes-boundary wavefunctions and, in general, different boundary
Hamiltonians (and different statistical ensembles in the Euclidean theory). In particular, in Sec. \ref{sec-microcanonical} we will discuss the microcanonical ensemble. Our goal here is not to classify all possible
choices, but to construct the doubled/joined system obtained from the
previously defined one-sided  theories. It will be an interesting future direction of research to analyze other boundary conditions.

\subsection{Tall geometries generated by matter in the presence of timelike boundaries}\label{sec:tall-example-geometries}

At least for global de Sitter, the Gao-Wald theorem ensures that matter satisfying {the null energy and null generic conditions} leads to a tall Penrose diagram with communication between the two poles\footnote{{The spacetime also needs to satisfy global hyperbolicity and null geodesic completeness.}}.  In principle, the problem requires re-analysis for the cases with timelike boundaries, since the solution space depends on boundary conditions (which are distinct from smoothness conditions required in the global case)\footnote{It is useful to reassess the assumptions that go into deriving the Gao-Wald theorem and how they manifest themselves in this timelike-bounded setup given its novel state space.  Regarding global hyperbolicity, given that we insist on a physically consistent solution of the initial boundary value problem, the future evolution is still determined from the initial conditions.
Null geodesic completeness physically holds, in that light rays bounce off the boundary consistently (rather than stopping there).
The relevant energy conditions and their degree of applicability in the physical (i.e. quantum) theory may need modification.  For example, the property of tallness is also sensitive to the Casimir energy due to the boundaries \cite{Batra:2024qju}, so the classical analysis is not enough to understand the space of tall or wide geometries in the spectrum.}.  For a discussion of this in two dimensions, also incorporating quantum effects, see \cite{Batra:2024qju}.    
In the present section, we simply show by example that the timelike boundaries do not remove the phenomenon of tall states.     

One example of this is provided by a homogeneous scalar field $\phi$ with Neumann boundary conditions at the Dirichlet (for gravity) walls.
Explicitly, consider the FLRW geometry
\begin{equation}
   ds^2 = -d\tau^2 +a(\tau)^2 ds^2_{S^3} = -d\tau^2 +a(\tau)^2 (d\theta^2 + \sin^2(\theta) ds^2_{S^2}) 
\end{equation}
with boundaries at $\theta=\theta_c$ {and $\theta=\pi-\theta_c$}
\begin{equation}\label{eq-boundary-global-coords2}
    \sin\theta_c = \frac{R_c}{a(\tau)}\,,
\end{equation}
consistently with the Dirichlet condition that the boundary be a Lorentzian cylinder of fixed proper radius $R_c$.
We prescribe a scalar field $\phi(\tau)$ which is homogeneous on the (bounded) $S^3$ directions and solves its equation of motion with Neumann boundary condition $\partial_\theta|_{\theta_c}\phi(\tau)=0$.  The Neumann boundary conditions ensure that the homogeneous field $\phi$ is a valid solution.   The tall feature can be seen from the Friedmann equation 
\begin{equation}
    \left(\frac{\dot a}{a}\right)^2 = \frac{\Lambda}{3} +\frac{\rho_\phi}{3 M_p^2} -\frac{1}{a^2}\,.
\end{equation}
For the case with $\rho_\phi=0$, the solution is $a(\tau)=\cosh(\tau/\ell)$ for de Sitter radius $\ell = \sqrt{3/\Lambda}$.  By the Gao-Wald theorem, this will be a tall geometry which is straightforward to derive explicitly; we see here that this is consistent with our Dirichlet conditions for gravity.

Another simple example (working in $2+1$D for simplicity) to illustrate the generalization of the Gao-Wald theorem to the system with boundaries and the causal connection between the two sides is  to consider two antipodal particles on the middle slice between the two boundaries, which source conical deficit angles, reducing the angular distance around them at a given proper distance from their position.  It is still possible to embed a boundary with fixed proper circumference $R_c$ in this spacetime.

For matter fields, the choice of boundary condition specifies the Hilbert space, related to the classical solution space; for Euclidean calculations it specifies the boundary ensemble. In this simple homogeneous scalar example we have imposed a Neumann condition so that the homogeneous scalar configuration is compatible with the timelike wall without turning on an additional boundary source. More general Dirichlet, Neumann, or Robin conditions are possible and would correspond to different Hilbert spaces and matter ensembles. Later in the paper, when discussing quantum matter and heat-kernel corrections, the boundary condition dependence will play an important role: quantum effects can renormalize the gravitational action and may motivate mixed or momentum-type gravitational boundary conditions.

\section{Joined timelike-bounded de Sitter geometry from a thermofield double:  maximum entropy and near-maximal mixing}\label{sec-top-joining-TFD-etc}
In this section, we begin the construction of our Hamiltonian theory of joined $\Lambda>0$ geometries,  first focusing on extending the geometry corresponding to the top band of energy levels dual to empty patches of de Sitter.

We take two copies of the finite quantum system \cite{Batra:2024kjl, Silverstein:2024xnr} corresponding to a given proper boundary radius $R_c$.  
We analyze the thermofield double (TFD) state of this system and its microcanonical analogue on both sides of the duality using the holographic dictionary, incorporating -- but generalizing to include essential matter and UV effects -- the construction introduced in \cite{Maldacena:2001kr} for a product of two CFTs\footnote{Reviewed pedagogically in \cite{hartman-lectures-2023}.}. This use of the dictionary is justified by the fact that the theory is a regulated multitrace deformation of gauge/gravity duality (with a finite real spectrum \cite{Coleman:2021nor, Batra:2024kjl, Silverstein:2024xnr}),  a procedure which preserves the essential framework of the duality dictionary.

To anticipate a little:  we will encounter various known subtleties in assessing the  consistency between the two sides of the duality, related to the fact that the de Sitter horizon is a maximal surface spatially \cite{Shaghoulian:2021cef} along with the fact that the de Sitter geometry itself is a maximum of the Euclidean gravitational action.
This will play out differently for the canonical and microcanonical TFD states.

For the microcanonical case it will be useful to work with the constrained path integral on the gravity side \cite{Banihashemi:2022jys}, which as we will see automatically removes such instabilities from the gravitational path integrand.  
For the canonical case, we will also work with the constrained path integral. Still, the timelike-bounded de Sitter geometry is a local maximum, rather than a minimum, of the constrained pure gravitational Euclidean path integral, which might suggest that the saddle point solution may not determine the gravity-side geometry or related quantities such as entanglement entropy.

A priori, the gravitational + matter path integral is not well understood to begin with.  Moreover, off-shell deformations of the Ryu-Takayanagi (RT) surface introduce curvature, rather than just conical singularities \cite{Lewkowycz:2013nqa}, calling into question the notion that the RT surface can controllably slip off the (spatially) maximal surface.
Still, it is interesting to pursue a more direct resolution in our framework.\footnote{In other frameworks, this is resolved in different ways. In the dS/dS framework \cite{Alishahiha:2004md, Alishahiha:2005dj, Dong:2010pm, Gorbenko:2018oov} the boundary theories live at the maximal slice, with no room to move the RT surface in between when we trace out one of the two identical sectors \cite{Dong:2018cuv}.  In the Cauchy Slice holography framework \cite{Soni:2024aop}, the Hamiltonian evolution along a Cauchy slice does not change the count of states.}  

First, we will show that in path integral calculations such as those relevant for computing the entropy, quantum matter effects can compete.  This effect is tied to the presence of the timelike boundaries:  the regime above the Hawking-Page transition entails a large spatial size of the boundary, of order the dS radius itself. In this regime, one loop QFT contributions can affect both the bulk geometry and the structure of the boundary value problem, leading us to an interesting new ensemble.   Moreover, it will be important to account for configurations in the path integral that involve the scalar uplifting from AdS to dS.  

Secondly, using the duality dictionary between the Brown-York and boundary energies, we will map the off-shell portion of the gravitational path integral to the integral over energies in the thermal partition function.  Our boundary theory implies crucial limits on this integral.  On one side of the unstable saddle, the range of integration is strongly bounded by the finiteness of our dressed spectrum \cite{Coleman:2021nor, Batra:2024kjl, Silverstein:2024xnr}.  On the other side of the saddle, it is limited by the sparseness of the spectrum below the band of energies corresponding to horizon microstates.  Altogether, this UV complete input removes the offending off-shell contributions, leaving intact the saddle point result for the entropy.

\subsection{Boundary theory setup: canonical thermofield double}\label{subsec:bdrytfd}

Let us start with the dual quantum mechanics (boundary theory) description of this state.  
Consider the doubled system prepared in a TFD state:
\be\label{eq-TFD}
|\psi_{\rm TFD}\rangle = {\mathcal N} \sum_i e^{-{\beta_c} \frac{E_i}{2}} |E_i \rangle |E_i \rangle\,,
\ee
where $i$ indexes the spectrum of
the one-sided finite quantum mechanics system \cite{Coleman:2021nor, Batra:2024kjl, Silverstein:2024xnr}, $E_i\le E_{max}$ and $\mathcal N$ is a normalization factor that we will determine shortly.  The entanglement properties of the state \eqref{eq-TFD} depend on the distribution of states $|E_i\rangle$ in the one-sided theory.  This theory has an exponentially large number $N_{max}$ of states at the top of its finite spectrum that make up most of the dimension of the Hilbert space with nearly degenerate energy $\sim E_{max}$.    

The remaining $N_{rest}=N-N_{max} \ll N_{max}$ states contribute subleadingly to the entropy. From the perspective of the deformation of AdS/CFT that leads to this theory, the $N_{max}$ states are deformed versions of the entropically dominant black hole mass that remain in the finite real spectrum, while the other states are lighter, including smaller black holes and EFT excitations.  {In our theory, these lower energy levels will be subject to a constraint defined below in \S\ref{sec-constraint-onto-joined-states}, which will encode the nontrivial causal connection between the two boundaries arising from the Gao-Wald theorem. As a result, the Hilbert space we will define does not in fact factorize altogether.  However, it approximately factorizes if we keep the most entropic band of energy levels unconstrained.  This entails keeping the individual microstates in the spectrum, a design choice for the Hilbert space we are constructing here.  Below in \eqref{eq-mctfd} we will introduce the microcanonical version of the thermofield double state in order to isolate the top band of energy levels, and in \S\ref{sec-constraint-onto-joined-states} we treat the rest of the states lower down in the spectrum.}

To connect to the discussions of maximal mixing \cite{Dong:2018cuv, Chandrasekaran:2022cip, Bousso:2000nf, Lin:2022nss, Banks:2000fe}, the normalization factor $ \mathcal N$ plays an important role. 
We can approximate the sum in the unit norm condition by the contribution of the entropically dominant states:
\be
1= | {\mathcal N}|^2 \sum_i e^{-{\beta_c} E_i} \approx | {\mathcal N}|^2 N_{max} e^{-{\beta_c} E_{max}}\; \Rightarrow\; {\mathcal N}= \frac{e^{{\beta_c} E_{max}/2}}{N_{max}^{1/2}}\,.
\ee
Then the TFD state is approximately
\be\label{eq-TFD2}
|\psi_{\rm TFD}\rangle \approx \frac{1}{N_{max}^{1/2}} \sum_a |E_a\rangle |E_a\rangle +\frac{1}{N_{max}^{1/2}} \sum_b e^{-{\beta_c} (E_b-E_{max})/2}\, |E_b \rangle |E_b \rangle
\ee
where the subindex $a$ stands for the entropically dominant states, and $b$ indexes the rest of the states.\footnote{
In order to quantify the error, let us write
\[
Z_1(\beta_c)=\sum_i e^{-\beta_c E_i}
= N_{\max}e^{-\beta_c E_{\max}}(1+\varepsilon_N),
\qquad
\varepsilon_N =
\frac{1}{N_{\max}}\sum_{b\notin{\rm top}}
e^{-\beta_c(E_b-E_{\max})}
+\varepsilon_{\rm band}.
\]
Here \(\varepsilon_{\rm band}\) denotes corrections due to the finite
width of the top band.  The approximation used above is valid when
\(\varepsilon_N\ll 1\), namely when the entropic enhancement of the
top band dominates the Boltzmann enhancement of lower-energy states and
when \(\beta_c\Delta E_{\rm top}\ll 1\) within the selected band.  In
that case
\[
{\cal N}
=
\frac{e^{\beta_c E_{\max}/2}}{\sqrt{N_{\max}}}
(1+\varepsilon_N)^{-1/2}
=
\frac{e^{\beta_c E_{\max}/2}}{\sqrt{N_{\max}}}
\left[1-\frac{\varepsilon_N}{2}+O(\varepsilon_N^2)\right].
\]
The associated corrections to the near-maximal entanglement are of the
same order.  In the microcanonical construction below the window
function restricts the sum to the top band, making this control explicit.}

We see from this that the entropically dominant states are in an approximately maximally mixed state, and effectively see an infinite temperature (called the ``Boltzmann temperature'' by \cite{Rahman:2024vyg}). But this is consistent with the fact that this state has a finite inverse temperature ${\beta_c}$.

This corresponds to the `tomperature' defined in \cite{Lin:2022nss} as being proportional to the variation of energy with respect to the number of qubits in the model.  We can apply this definition explicitly to the dressed energy for entropically dominant states in \cite{Coleman:2021nor, Batra:2024kjl, Silverstein:2024xnr} as follows.  For boundary space-time dimension $d$, the dressed energy is
\be\label{eq:dressed-gen}
{\cal E}_n(y)=\frac{d(d-1)}{2\pi y} \left[\sqrt{1+ \Omega^{2/(d-1)} (C_d y)^{2/d}}\Big|_{y^{ (\lceil d-2 \rceil)/d }} \mp \sqrt{\eta+\Omega^{2/(d-1)} (C_d y)^{2/d}-\frac{4\pi}{d(d-1)}y{\cal E}_n(0)}\; \right]\,,
\ee
along with matter corrections that do not change the count of states or top-heavy structure of the spectrum, and at least in some cases simply multiply the second square root \cite{Batra:2024kjl, Silverstein:2024xnr}.  Here the first square root is a formal expression, expanded out to the indicated order as explained in \cite{Silverstein:2024xnr}. It corresponds to a specific choice of local terms at the boundary in the gravitational action; other choices are also possible.  
$\Omega=2\pi^{\frac{d-1}{2}}/\Gamma\left(\frac{d-1}{2} \right)$ is the volume of the unit radius $(d-1)$--dimensional sphere, $y=\lambda/V^{d/(d-1)}$, $V= \Omega R^{d-1}$, and ${\cal E}= V^{1/(d-1)} E$. The relation between holographic and gravitational parameters is
\be
\frac{\lambda}{V^{d/(d-1)}} = \frac{4d G_N \ell}{V_c^{d/(d-1)}}\,,
\ee
and  
\be\label{eq:translation}
C_d =\frac{1}{4d}\frac{\ell^{d-1}}{G_N}\;,
\ee
is the count of degrees of freedom of the seed theory (e.g. given by the seed CFT central charge in the bulk $d+1=3$ dimensional case). { The energy $E$ corresponds to the Brown-York energy \cite{Brown:1992br} in Lorentzian signature,
\be \label{eq-BY-tensor}
T_{\mu\nu}= -\frac{1}{8\pi G_N} (K_{\mu\nu}-K h_{\mu\nu}+ \ldots)\;,\;E_{BY}=- \int d^{d-1} \bar x \sqrt{h^{(d-1)} }\,T^0_0\,.
\ee
$K_{\mu\nu}$ is the extrinsic curvature of the $d$-dimensional time-like boundary, $h_{\mu\nu}$ is its induced metric, and $d^{d-1} \bar x \sqrt{h^{(d-1)} }$ is its spatial volume element. The `$\ldots$' are local counterterms that can be chosen  freely for the finite boundary; one choice is as in 
(\ref{eq:dressed-gen}), as discussed in \cite{Silverstein:2024xnr}.}
For $d=2$ we have
\begin{equation}\label{eq-2d-pure-gr-energy}
  {\cal E}_n= E_nL = \frac{1}{\pi y}\left(1\mp \sqrt{\eta +\frac{y}{y_0}(1-\eta)-2\pi y {\cal E}_n^{(0)}  + 4\pi^4 J^2 y^2\,}\right)\,,  ~~~~ (\text{pure}~ T\bar T+\Lambda_2),
\end{equation}
where ${\cal E}_n^{(0)}=2\pi (\Delta_n-c/12)$ is the dimensionless energy of the $n$th level in the seed CFT, with $\Delta=c/6$ for the top energy band corresponding to empty (matter-free) de Sitter. 

We can now connect to the proposal in \cite{Lin:2022nss} that the finite notion of temperature -- dubbed `tomperature' -- that corresponds to the static patch observer-measured thermal particle spectrum can be understood as the change in the energy upon changing the number of qubits of the system.  Our setup generalizes this to the timelike-bounded system, rather than just an observer situated at the pole.  Working with the top energy band and identifying the number of qubits with the central charge $c$, we find
\begin{equation}
\frac{\partial}{\partial c} {{\cal E}}\propto 1/\sqrt{(y\pi^2 c/3)-1} = 1/\beta_c
\end{equation}
where $\beta_c$ is the boundary Euclidean time circle circumference in the Euclidean version of the $T\bar T + \Lambda_2 +\dots$ deformed theory, using the duality dictionary derived in \cite{Coleman:2021nor}.
This accords with the first law \cite{Banihashemi:2022htw} $T_c \, dS=dE_{\rm dressed}$.  

In the next subsection, we will construct the gravity-side avatar of this state, finding that it corresponds to a joined timelike-bounded de Sitter geometry in its Hartle-Hawking state, analogously to the Hartle-Hawking state in the two-sided AdS black hole in \cite{Maldacena:2001kr}. This calculation will lead to a geometry for which the inferred inverse temperature $\beta_c$
corresponds to the expected gravity side quantity.
That is, the value of $\beta_c$ on the gravity side matches the standard notions of temperature for a timelike-bounded holographic system; this is usually known as the `local' or `Tolman temperature.'  It depends  on the boundary proper size as in \cite{Coleman:2021nor} (and similarly in four dimensions \cite{Silverstein:2024xnr}).  It can be determined as the circumference 
of the compact Euclidean time direction of a smooth Euclidean continuation of timelike bounded empty de Sitter, Eq.~(\ref{eq:emptydS}).  It can be understood directly in Lorentzian signature as the temperature measured by an Unruh detector following a trajectory at fixed spatial position on the boundary. It can also be obtained from other probe calculations, such as from the periodicity of 2-point functions of light operators in the gravity background.

\subsection{Gravity side: state preparation for timelike-bounded dS canonical thermofield double and its UV sensitivity}\label{subsec:preparation}

Let us analyze the Hartle-Hawking prescription for a thermofield double at tomperature $\beta_c$.     
To start, we recall that in the AdS/CFT seed theory, the state \eqref{eq-TFD} can be constructed by a Euclidean path integral in the boundary theory on a $S^{d-1}\times I$ of sphere radius $R_c$ and interval length $\beta_c/2$. This prescription extends to the deformed theory as well:   as described in \cite{Batra:2024kjl}, the finite Hamiltonian system can be formulated using a path integral, with the finite spectrum emerging from it via the square root in the path integral phase.

Given this, we next apply the holographic dictionary, analyzing the gravitational path integral with Dirichlet boundary conditions enforcing that the boundary be $S^{d-1}_{R_c}\times I_{\beta_c/2}$.
The gravitational path integral is not generally well defined, with no known a priori derivation of its rules concerning the appropriate contour prescription and order of integration required to match the right physics, although there is much insightful work on the topic; see e.g., \cite{Halliwell:1988ik,Halliwell:1989dy, Colin-Ellerin:2020mva, Marolf:2022ybi, Dittrich:2024awu, Held:2026huj,Chou:2024sgk,Kolanowski:2026gii}. 
Many studies above two dimensions focus on semi-classical solutions.  In the de Sitter context, as we will review shortly, the classical saddles for the partition function and for the related calculation of the entanglement entropy are local maxima of the Euclidean action. Despite being maxima, we will find that these saddles give sensible results for various quantities. Even before getting into important quantum and microphysical effects which will impact the putative role of off-shell contributions, it is not a priori obvious that the maximum is physically subdominant, as a simple example like the following illustrates. Perhaps a useful analogy to keep in mind is that of a massive particle in a Euclidean sphere, suggested in Sec. 3 of \cite{Maldacena:2024spf}. The classical saddle (corresponding to a particle going around a great circle of the sphere), is a maximum but nevertheless can provide a good approximation to the full partition function.\footnote{A key difference is that in the gravitational case so far we lack a well-defined way of performing the gravitational path integral.}  We will now analyze our case systematically.  

\begin{figure}[h]
    \centering
    \includegraphics[width=0.4\linewidth]{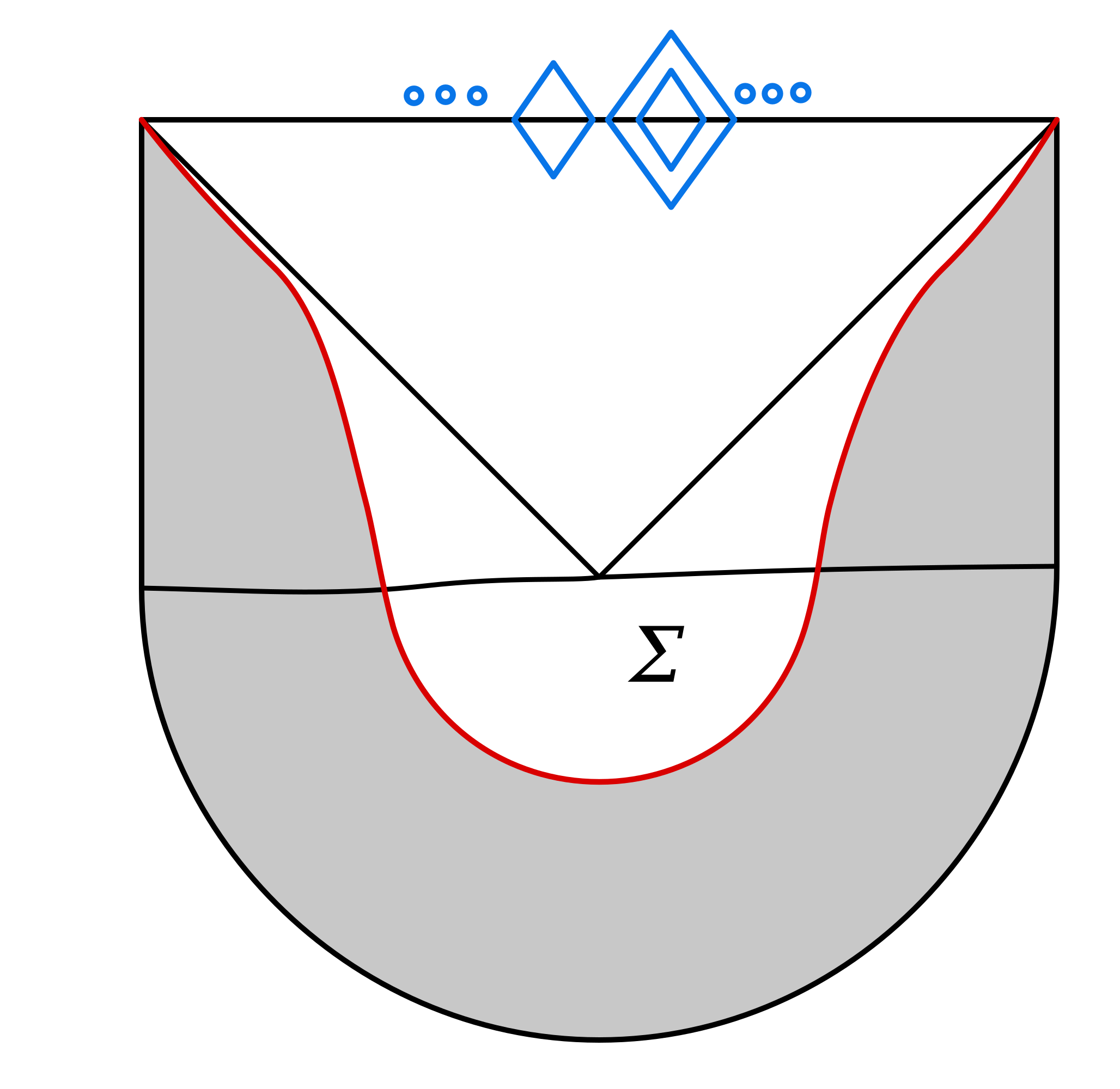}
    \caption{Above the Hawking-Page transition, the dominant saddle has a large boundary and satisfies $\beta_c<L_c$,where $\beta_c$ is the boundary inverse temperature and $L_c$ is the boundary circumference. The shaded region is {\it excluded} in this figure; the existing spacetime (in white) joins the two large boundaries on the spatial slice $\Sigma$. At the level of the purely gravitational path integral, this saddle---and that of the closely related calculation of the partition function---is unstable:  changing the spatial size $L_0$ at the $Z_2$-symmetric central slice of the geometry reduces the Euclidean action (cf \eqref{I-can-L0} in bulk dimension $d+1=3$).  However, its microcanonical counterpart is classically stable as shown in \S\ref{sec-microcanonical}.  At the quantum level with matter,
    as discussed in \S\ref{sec-UVsensitive}-\ref{sec-UV-QFT-illustration} this system is UV sensitive, subject to bulk QFT corrections.  At the non-perturbative level, the microscopic dual formulation implies limitations on the off-shell gravitational path integral, restricting the range of integration of $L_0$ (\S\ref{subsec:UVcomplete}).} 
    \label{fig:TFD_bdry}
\end{figure}

First, consider the classical on-shell infilling of the bulk geometry.  In a fully Euclidean solution, this simply entails cutting out a ball 
with boundary $S^{d-1}$ surrounding a great circle on the sphere.  The relevant portion of this geometry is half of it: the slice $\Sigma$ at the boundary of a hemisphere provides the initial spatial slice of the timelike bounded Lorentzian geometry (see figure \ref{fig:TFD_bdry}). This on-shell geometry gives a saddle point approximation to the Euclidean gravity path integral, with classical action
\be\label{eq:Ifull}
I= I_{\rm bulk}+I_{\rm GHY}\,,
\ee
where
\be
I_{\rm bulk}= - \frac{1}{16\pi G_N} \int_M d^{d+1}x\,\sqrt{g} \left[R^{(d+1)}- \frac{d(d-1)}{\ell^2} \right]\,,
\ee
and
\be
I_{\rm GHY}=- \frac{1}{8\pi G_N} \int_{\partial M} d^dx\,\sqrt{h} \left[K+ \ldots\right]\,.
\ee
The `$\ldots$' are additional local geometric terms that won't be important in what follows. When comparing boundary and bulk quantities we will commit to the gravitational counterterms that produce the first term in (\ref{eq:dressed-gen}).

The Euclidean sphere
\be\label{eq:sphere}
ds^2= f(r) d\tau^2+ \frac{dr^2}{f(r)}+r^2 d\Omega_{d-2}^2\;,\;f(r) = 1- \frac{r^2}{\ell^2}
\ee
then gives two on-shell solutions in the presence of a Dirichlet wall at $r=R_c$: the cosmic horizon (CH)  patch when $R_c \le r \le \ell$, or the pole patch (PP) for $0\le r \le R_c$. Regularity in the cosmic horizon patch dictates the periodicity $\tau \to \tau+ 2\pi \ell$, while regularity in the pole patch requires $\phi \to \phi+2\pi$. So this particular geometry corresponds to 
\be\label{eq:betac1}
\beta_c=2\pi \ell \sqrt{1-R_c^2/\ell^2}\,.
\ee
We will consider below more general on-shell geometries corresponding to dS black holes, where $\beta_c$ can be chosen independent of $R_c$ and $\ell$. 

Among solutions to the bulk gravitational equations of motion (without matter), the cosmic horizon patch dominates only above the (de Sitter) Hawking-Page (HP) transition.
To derive this, we substitute the above metric into the Euclidean action, finding\footnote{The leading additional counterterms not shown in $I_{\rm GHY}$ above, which are the wall cosmological constant and Ricci scalar terms, cancel out from $I_{\rm CH}-I_{\rm PP}$, so they do not affect the HP transition calculation.}
\be
I_{\rm CH}-I_{\rm PP}=\frac{\textrm{Vol}(S^{d-1}) \ell^{d-1}}{4 G_N} \left[2(d-1) \left(1-\frac{R_c^2}{\ell^2}\right) \left(\frac{R_c}{\ell}\right)^{d-2}-1 \right]\,.
\ee
Therefore, the CH patch dominates if
\be\label{eq:HPbound}
\frac{d-1}{2\pi^2} \left( \frac{\beta_c}{\ell}\right)^2  \left(\frac{R_c}{\ell}\right)^{d-2}<1\,.
\ee
This HP transition was observed for $d=2$ in \cite{Coleman:2021nor}.

\subsubsection{Constrained off-shell analysis}
\label{sec-Constrained-off-shell}

It turns out that the geometry (\ref{eq:sphere}) does not have minimal classical Euclidean action -- there are off-shell configurations whose Euclidean path integrand is larger.  In these configurations, the central slice has an area different than the horizon size in empty de Sitter.
Let us derive this in the case of three dimensional de Sitter, in a similar way as the higher-dimensional analysis in \cite{Banihashemi:2022jys}. In particular we work out the Euclidean action for the geometry appropriate to calculating the partition function, which starts from a boundary containing a compact circle of size $\beta_c$ and then fills in the bulk geometry.    

We consider geometries\footnote{For a generalization to $\tau$-dependent geometries, see appendix \ref{t-dependence}.} 
\be \label{U1-sym-metric}
ds^2=b^2(r) d\tau^2 + dr^2+ L^2(r) d\phi^2/4\pi^2\, , \quad \tau \equiv \tau + P_\tau \, , \, \phi \equiv \phi + P_\phi \, , 
\ee
with periodic coordinates $\tau$ and $\phi$. With coordinate rescalings we may always pick $P_\tau =P_\phi=2\pi$. 
The coordinate $r$ parameterizes the radial direction, and we take its range to be $r \in [0,r_c]$, with $r_c$ indicating the location of the boundary. The boundary conditions are
\be
2\pi  b(r_c) = \beta_c \quad , \quad  L(r_c)=L_c \, ,
\ee
so we fix $(\beta_c, L_c)$ at the system boundary. 

There are two categories of configurations depending on whether the $\tau$-circle shrinks to zero size at $r=0$ or not.\footnote{The geometry would not be smooth if the $\tau$-circle closes off somewhere between $[0,r_c]$ and reopens, so we don't consider such cases.} If it closes off, we have $b(0)=0$ and the manifold ends at a ``horizon'' (or the set of fixed points of $\tau$-translation). If it doesn't, there would be another boundary at $r=0$ unless $L(0)=0$. Thus for a single-boundary system we either have $b(0)=0$ or $L(0)=0$. In the presence of a horizon, the regularity condition at the center is
\be \label{reg-b}
 b'(0)=1 \,, \quad  {\rm when} \, \,b(0)=0\,.
\ee
The Euclidean action with the Gibbons-Hawking-York (GHY) boundary term is given by (\ref{eq:Ifull}) for $d=2$, with cosmological constant $\Lambda=1/\ell^2$.
The Ricci scalar for the above metric is
\be
\label{eq:scalarC}
R= -\frac{2}{bL}\left( b'L'+bL''+b''L \right )\,.
\ee
After some integrations by parts, we obtain
\be \label{I-bL}
I=\frac{1}{4 G_N}\left [ \int_0^{r_c}\!dr\, b(L''+\Lambda L)-(Lb')_0-(bL')_{r_c}\right].
\ee
The boundary term at $r=0$ in \eqref{I-bL} vanishes for the case without a horizon (when $L(0)=0$). For a configuration where the $\tau$-circle closes off smoothly, we have
\be \label{off-shell-action}
I=\frac{1}{4 G_N} \int_0^{r_c}\!dr\, b(L''+\Lambda L)-\frac{L_0}{4G_N}-\frac{\beta_c}{8\pi G_N} L'(r_c)\, ,
\ee
where we used the boundary condition for $\beta_c$ and introduced
\be
L_0 :=  L(0) \, .
\ee
The action \eqref{off-shell-action} may be used for both cases, by setting $L_0=0$ when there is no horizon. For fully on-shell configurations the last term becomes $\beta_c$ times the Brown-York energy $E_{BY}$.  

The bulk radial integral in the Euclidean action is in general unknown. However, we note that the integrand $L''+\Lambda L$ vanishes if we only impose the Hamiltonian constraint ($\tau$-translation generator).\footnote{Note that for the class of metrics \eqref{U1-sym-metric} the momentum constraint is satisfied due to $\tau$-translation symmetry.} The justification for imposing the constraints is provided by including only the {\it physical} states in the path integral; see e.g., \cite{Whiting:1988qr}. Imposing the Hamiltonian constraint and requiring a regular geometry we find
\be
L(r)=L_0 \cos  r/\ell \, ,
\quad \quad \text{(constrained configurations)} 
\ee 
with $L_0$ indicating the ``area'' of the codimension-2 surface where the $\tau$-circle shrinks to zero size, and $\ell=1/\sqrt{\Lambda}$ being the dS$_3$ length. Thus the value of $r_c$ is related to the boundary condition by $L_c=L_0 \cos r_c/\ell$. We note that the value of $L_0$ is larger than $L_c$.

\begin{figure}[h]
    \centering \includegraphics[width=0.7\linewidth]{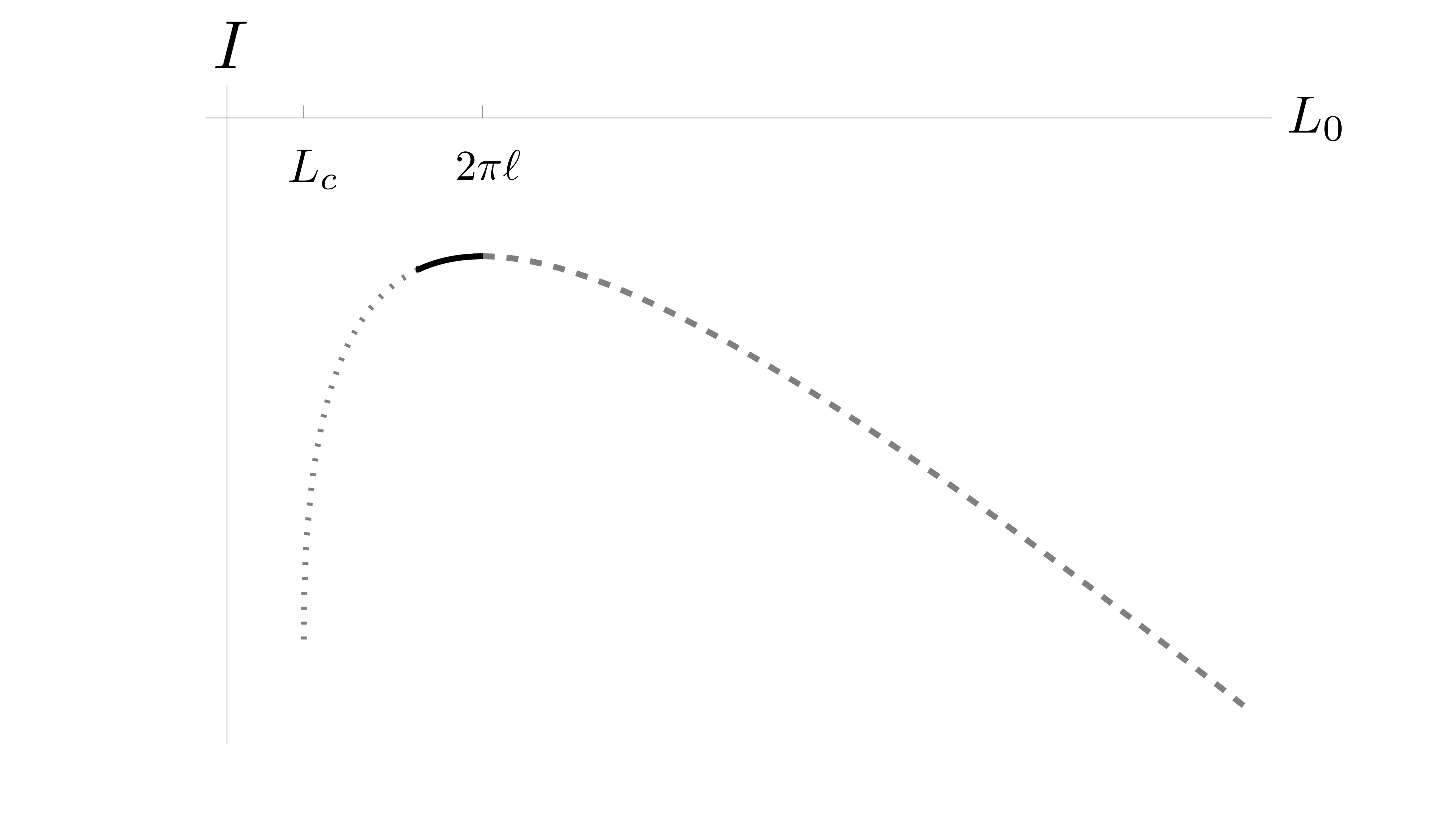}
    \caption{Canonical Euclidean action \eqref{I-can-L0} in terms of $L_0$ for a given $L_c$ above the HP transition \eqref{eq:HPbound} and temperature \eqref{on-shell-beta} corresponding to the pure dS solution with $L_h=2\pi \ell$. The dashed segment with $L_0>2\pi \ell$ indicates geometries with larger boundary energy (sourced by a conical excess in the nonexistent region behind the boundary, as if from a negative mass there).  These states are absent in our UV complete finite spectrum, derived by keeping the real spectrum along the $T^2 + \Lambda + \dots$ deformation \cite{Coleman:2021nor, Batra:2024kjl, Silverstein:2024xnr}.   The dotted segment on the left represents configurations that are also absent, in that they correspond to energies which are very sparsely populated in the spectrum of our UV complete deformed theory. The remaining black segment depicts a single highly populated band of energies corresponding to the de Sitter microstates. See \S\ref{subsec:UVcomplete} for further details.
    }
    \label{fig-canonical-action}
\end{figure}

The action becomes
\be \label{I-can-L0}
I[\beta_c, L_c; L_0]= -\frac{L_0}{4G_N}+\frac{\beta_c}{8\pi G_N \ell}\sqrt{L_0^2-L_c^2}\, .
\ee
This can be viewed as a ``constrained" action, which is partially off-shell and depends on the variable $L_0$. The on-shell value of this metric degree of freedom $L_0$, which we label by $L_h$, can be obtained by solving $\partial I/\partial L_0|_{L_h}=0$. This relates the on-shell value $L_h$ to the ensemble parameters $(\beta_c,L_c)$ via 
\be \label{on-shell-beta}
\beta_c =2\pi \ell \sqrt{1-L_c^2/L_h^2} \quad \quad \quad \text{(on-shell).}
\ee
It is easy to see that for a generic set of boundary values $(\beta_c,L_c)$ the action \eqref{I-can-L0} is a local maximum as a function of $L_0$ at the stationary point $L_h$, and is unbounded below for arbitrarily large $L_0$; see figure \ref{fig-canonical-action}.
So there are off-shell configurations that are smooth, but have lower Euclidean action compared to the on-shell one. Smoothness can be guaranteed by judiciously picking an appropriate $b(r)$ that satisfies all the boundary and regularity conditions, but not the full set of field equations. 

Similarly, in a calculation of the Ryu-Takayanagi surface obtained from a partition function with boundary circle of size $n\beta_c$ \cite{Lewkowycz:2013nqa}, the `cosmic brane' \cite{Dong:2016fnf} associated with the RT surface appears to cost less action when it shrinks off of the maximal surface at the horizon.
This so far might naively suggest that the entanglement entropy between our two systems (each serving as the other's heat bath) is reduced compared to the horizon area in Planck units.  

\subsubsection{UV sensitivity}\label{sec-UVsensitive}

But the situation is more subtle: the regime above the Hawking-Page transition is UV sensitive and will not in general be governed by the classical general relativistic dynamics.  
First, for the RT calculation just mentioned, we note that as in \cite{Lewkowycz:2013nqa, Dong:2016fnf}, if we move the conical singularity off of the extremal surface, it worsens into a curvature singularity.
The argument uses the locally Gaussian coordinates around the Renyi brane (fixed point of the replica symmetry) that appears in the orbifolded picture ${\mathcal M_n}/\mathbb Z_n$, where $\mathcal M_n$ is the replicated manifold, and $\mathbb Z_n$ is the replica symmetry. The metric reads
\be
ds^2_{local}=n^2 dr^2+ r^2 dt_E^2+ \gamma_{ij} dy^i dy^j\,,
\ee
where the transverse metric admits an expansion
\be
\gamma_{ij}=\gamma_{ij}^0+ 2 K^{(1)}_{ij} r^n \,\cos t_E+ 2 K^{(2)}_{ij} r^n \,\sin t_E+ \ldots
\ee
Here $K^{(1)}$ and $K^{(2)}$ are the two extrinsic curvatures of the Renyi brane.
A short calculation shows that the Ricci tensor becomes proportional to $(n-1) K^{(a)}/r^{2-n}$ as we approach the defect; see also \cite{Dong:2016hjy}. Therefore, if the defect is not extremal (namely if the extrinsic curvatures are non-vanishing), the backreacted geometry develops curvature singularities. These singularities will also appear in the covering space $\mathcal M_n$, which locally amounts to increasing the periodicity of $t_E$ by a factor of $n$. We conclude that such nonextremal configurations will generically not give rise to a nonsingular geometry, calling into question its contribution to the path integral calculation of the Renyi and entanglement entropies.

Even in the calculation of the partition function (which does not involve the conical defect), there is UV sensitivity.  We will start by exhibiting this via a simple estimate of back reaction, and then detail its effects in \S\ref{sec-UV-QFT-illustration}.  
Consider the boundary layer of quantum field stress energy that accumulates, as derived in \cite{Deutsch:1978sc} (with phenomenological consequences recently analyzed in \cite{Philcox:2025faf}).  We can appraise the effect of this on the geometry by estimating the curvature sourced by our boundary layer of stress energy $T_{\rm QFT}$, an energy density of order $M_{UV}^4$ extending a thickness of order $1/M_{UV}$ from the boundary.    Approximating the Greens function by $1/\Delta X^2$ for a point $x$ a distance $\sim \Delta X$ from the boundary, and integrating over source points on the boundary within a distance of order $\Delta X$ from the boundary yields the scaling   
\begin{eqnarray}\label{eq-UV-sensitivity-estimate}
\delta R_{\mathrm{QFT}}(x) &\sim& G_N \, \partial_x^2 \int d^4 x' \, G(x, x') \, T_{\mathrm{QFT}}(x') \nonumber \\ 
&\sim& \frac{G_N}{\Delta X^4} \left( M_{UV}^{-1} \, \Delta X^3 \right) M_{UV}^4 \sim \frac{1}{\ell_{dS}^2} \left[ \left( G_N \, M_{UV}^2 \right) \frac{\ell_{dS}^2 \, M_{UV}}{\Delta X} \right] \,.  
\end{eqnarray}
The last factor in square brackets is much larger than one; if we put $x$ on the central slice in between the two boundaries, this is of order $\ell_{dS}M_{UV}$, and otherwise it is larger. Thus it is controlled by the four dimensional Hubble scale in units of the fundamental cutoff $M_{UV}$.  The factor $G_N M_{UV}^2$ on the other hand is less than one, but it is controlled by a factor of the internal volume of the string compactification in the same fundamental cutoff units \cite{Silverstein:2004id, Silverstein:2016ggb}.  This internal scale yields a much smaller hierarchy than that involving the Hubble scale in standard quantum gravitational de Sitter models.  

As a result, the curvature sourced by the quantum stress energy accumulated on the boundary in the relevant regime above the Hawking-Page transition (where the boundary is larger than $\beta_c$) could quite easily compete with the background curvature.  This raises the interesting possibility that the canonical TFD -- along with derived quantities such as the entanglement entropy obtained by tracing over one system --  is stabilized by quantum matter, analogously to the stabilization of dS itself through quantum or stringy effects.

\subsubsection{Illustration that matter QFT effects can affect entropy calculations}\label{sec-UV-QFT-illustration}

To elaborate on this, consider the full partition function including matter.  Schematically, for example for a scalar field, we have 
\begin{flalign}\label{eq:PIwithscalar}
    Z[\beta,L_c] &= \int \mathcal{D}g \,\mathcal{D} \phi \,e^{-I_0[g]-I^{\text{scalar}}[g,\phi]} \; \\ 
    I^{\text{scalar}}[g,\phi] &= \frac12 \int d^3 x\sqrt{g}  \, \left(g^{\mu\nu}\partial_\mu \phi \partial_\nu \phi+m^2\phi^2+\xi R\phi^2 + \text{interactions}\right)  \;, \nonumber
\end{flalign}
where $I_0$ denotes the pure gravitational Euclidean action.

\noindent \textbf{Gaussian scalar.}  Integrating out a Gaussian scalar field already introduces nontrivial dependencies on the bulk and boundary gravitational sector (see e.g. \cite{Vassilevich_2003} and references therein for a summary).  We need to know the effect of quantum matter on the Euclidean action, with an eye toward how it affects the $L_0$ dependence.  
We will incorporate general possibilities for the scalar boundary conditions.  The result for the renormalized action was computed in general in \cite{Vassilevich_2003} using the heat kernel method, in terms of a small regulator $\epsilon$ with dimensions of length.  


There are many invariant contributions to the bulk and boundary action, including a renormalization of Newton's constant 
along with other terms that generally depend on the boundary condition on the scalar.  We can consider a class of (generally inhomogeneous) Robin conditions
\begin{align}\label{eq:inhomoRobin}
    n^\mu \partial_\mu \phi+S(x)\,  \phi |_{\partial M}= j(x) \;. 
\end{align}
To implement this, we need to add an extra boundary term $  \int_{\partial M} d^2 x\sqrt{h} \left(\frac12 S \phi^2- j\phi \right)$. 
We use the standard heat-kernel expansion
\be
K(t)=\mathrm{Tr}\, e^{-t(-\Delta)}
\simeq \frac{1}{(4\pi t)^{D/2}}
\sum_{n=0}^{\infty} a_n\, t^{n/2}\,,
\ee
where the $a_n$ are local bulk and boundary invariants determined by
the field content and boundary conditions, and $\Delta$ is the quadratic kernel determined from $I^{\textrm{scalar}}$ above.
The form of each term is completely fixed by dimensional analysis, and the contributions that scale like $1/\epsilon^3$,  $1/\epsilon^2, 1/\epsilon,$ and $\log(\epsilon)$ respectively are \cite{Vassilevich_2003}
\begin{align}\label{eq:HKcoe3D}
a_0 & \propto \int_M d^3 x\sqrt{g}\nonumber\\
    a_1 & \propto \int_{\partial M} d^2 x\sqrt{h}\nonumber\\
    a_2 & \propto \int_M d^3 x\sqrt{g}\left( R-6m^2\right)+\int_{\partial M} d^2 x\sqrt{h} \left( b_1 K + b_2 S\right)
    \nonumber\\
    a_3 & \propto \int_{\partial M} d^2x\,\sqrt{h}\;
\left[
c_1 m^2
+ c_2 R
+ c_3 R_{nn}
+ c_4 K^2
+ c_5 K_{ij}K^{ij}
+ c_6 S K
+ c_7 S^2
\right] \; .
\end{align}
In these expressions, $S$ is the function appearing in the Robin condition \eqref{eq:inhomoRobin}. If we consider say Dirichlet, there won't be terms involving $S$. We have denoted $R_{nn}=n^\mu n^\nu R_{\mu\nu}$, and have used the fact that in 3D the Riemann tensor can be written in terms of the Ricci tensor and the metric (in higher dimensions, there will be terms involving $e^\mu_i e^\nu_j n^\alpha n^\beta R_{\mu \alpha \nu \beta}$).

Interestingly, these contributions can include a term of order $L_0^2$ in the action, supplementing the classical terms $\sim L_0, \sqrt{L_0^2-1}$.  The contribution to the boundary action $\sim S^2$, with $S$ tied to the trace of the extrinsic curvature, $K$, could give a contribution that is positive.  Perturbing around our original configuration, this would produce a term of order $L_0^2$ meaning we have classical + quantum dependence on $L_0$ of the form
\begin{equation} \label{3term-action}
    I_E \sim -\frac{L_0}{4G_N}+\frac{\beta}{8\pi G_N}\sqrt{L_0^2-L_c^2} + \eta \Lambda L_0^2 +\dots
\end{equation}
showing that it can in principle affect the conclusions about the location and stability of the $L_0$ saddle; new dependence on $b(r)$ also arises.   However, this affects the variational problem for gravity \cite{Jacobson:2013yqa}, requiring a detailed back reacted analysis as a function of model parameters including field content and boundary conditions. Sticking to the Dirichlet condition to define the full path integral, in other words, can force the configurations to be off-shell.  A complementary approach in such cases is to work with a different boundary condition which is compatible with on-shell solutions to the matter-corrected effective action.  See \S \ref{sec-new-bc} for an example of a novel boundary condition we discovered in this way, which stabilizes the $L_0$ direction in the path integral.

Moreover, interactions are important in top down de Sitter uplifts of AdS/CFT \cite{Dong:2010pm, DeLuca:2021pej}, requiring analysis beyond the Gaussian approximation. 
We turn to a particular structure of interactions involved in the uplift sector next.

\noindent \textbf{Uplift sector matter effects.}
Top-down de Sitter models useful for holography have been constructed as uplifts of known AdS/CFT models \cite{Dong:2010pm, DeLuca:2021pej}.  From these we can abstract a basic structure -- an uplift potential -- that will be part of the scalar field sector \eqref{eq:PIwithscalar}.   The uplift structure enters into the single-sided theories obtained by deformation of AdS/CFT \cite{Batra:2024kjl, Silverstein:2024xnr}, so it is intrinsic to our holographic models.  

Let us define an uplift scalar $\phi_u$ as in \cite{Batra:2024kjl, Silverstein:2024xnr}, which has a potential $V_u(\phi_u)$ including an AdS minimum at $\phi_u=\phi_A$ and a dS local minimum at $\phi_u=\phi_S$.  This yields a contribution
\begin{equation}\label{eq:uplift-scalar-action}
    I_u =\frac{1}{2} \int d^{d+1}x \sqrt{g}\left\{(\nabla \phi_u)^2 + V_u(\phi_u)\right\}\,
\end{equation}
to the Euclidean action.
Within a path integral calculation of thermodynamic quantities, this contribution of $I_u$ to the Euclidean action could potentially be reduced by exploring configurations that access the AdS (or perhaps also asymptotically flat) regions of the potential landscape.  This costs gradient energy, but gains in potential energy, relative to the fiducial configuration $\phi_u=\phi_S$.

We can obtain a rough estimate for the various contributions to the Euclidean action as follows.   
Modeling the configuration as a thick wall with $\partial_r\phi \sim \phi/\ell$, we can write the gradient term in $I_u$ as \be
I_{gradient}\sim (\frac{\Delta\phi_u}{M_p})^2 (\frac{M_p}{\ell})^2 \,{Vol}
\ee
where ${Vol} = {Vol}_{dS}+{Vol}_{AdS}$ is the volume between the boundary and the horizon, decomposed into pieces which are approximately dS or AdS.  A similarly rough estimate of the potential term is 
\be    
I_V \sim \frac{-M_{p}^2}{\ell_{AdS}^2} (Vol_{AdS}) + \frac{M_{p}^2}{\ell_{dS}^2} (Vol_{dS}).
\ee
The net gain for $\ell_{AdS}=\ell_{dS}$ is approximately
\begin{align*}
    \Delta I_{Eucl} &\approx \frac{M_p^2}{\ell^2} \left[ -2 Vol_{AdS} + Vol_{dS} + \left(\frac{\Delta\phi_u}{M_p}\right)^2 (Vol_{AdS} + Vol_{dS}) \right] \\
    &= \frac{M_p^2}{\ell^2} \left[ -Vol_{AdS} \left(2 - \left(\frac{\Delta\phi_u}{M_p}\right)^2\right) + Vol_{dS} \left(1 + \left(\frac{\Delta\phi_u}{M_p}\right)^2\right) \right]\,.
\end{align*}
For $\Delta\phi_u < 2 M_p$, this can be made $< 0$ if:
\begin{equation*}
    Vol_{dS} < Vol_{AdS} \left( \frac{2 - \left(\frac{\Delta\phi_u}{M_p}\right)^2}{1 + \left(\frac{\Delta\phi_u}{M_p}\right)^2} \right)
\end{equation*}
providing a net gain (lower Euclidean action, dominating over the pure-dS saddle).  The domain wall introduces a region near the horizon for which the horizon is a spatial minimum, removing the pure dS feature that it is a maximum.

\subsection{Back to the Boundary: stabilizing the canonical partition function} \label{subsec:UVcomplete}

Let us now incorporate into the gravitational path integral essential features of our UV complete boundary theory description.  We will show how this resolves the apparent instability in $L_0$ \eqref{I-can-L0}, by relating $L_0$ to the Brown-York energy, and inputting the basic structure of our dressed energy spectrum \cite{Coleman:2021nor, Batra:2024kjl, Silverstein:2024xnr} which is sparse below a single highly populated band of energies corresponding to the de Sitter microstates.
For simplicity we will present the formulas for $d+1=3$, but similar considerations apply in higher dimensions in terms of the $(TT)_s$ deformation of \cite{Silverstein:2024xnr}. 

First, recall that the  one sided de Sitter microstates at the top energy band, which dominate the entropy, are obtained by dressing the AdS black hole microstates near the Hawking-Page transition \cite{Coleman:2021nor, Silverstein:2024xnr}. The result for this band of dressed energies -- first expressing this at the level of pure gravity in the bulk -- in $d+1=3$ is
\begin{equation}\label{eq:2d-pure-gr-energy}
  {\cal E}_n= E_nL = \frac{1}{\pi y}\left(1+ \sqrt{\eta +\frac{y}{y_0}(1-\eta)-2\pi y {\cal E}_n^{(0)}  + 4\pi^4 J^2 y^2\,}\right)\,,  ~~~~ (\text{pure}~ T\bar T+\Lambda_2),
\end{equation}
where $\eta=-1$ in the de Sitter part of the trajectory { and $y=\lambda/L^2$ is the dimensionless deformation parameter}. The matching point occurs at the $T\bar T$ coupling
\be
y_0= \frac{1}{4\pi^2 (c/12+\delta)}
\ee
with large central charge $c$.  The parameter $\delta$ here is related to the width of the top band of energy levels around the Hawking-Page transitions, as we describe below; as such we may define $y_0$ anywhere in this band.
Furthermore, the undeformed energies are related to the CFT scaling dimensions by
\be
{\cal E}_n^{(0)}=2\pi \left(\Delta_n- \frac{c}{12} \right)\,,
\ee
and $J$ is the angular momentum, which for simplicity in the following we set to zero (states with nonvanishing angular momentum are entropically subdominant). Incorporating matter fields (and gravitons in the higher dimensional case) retains the count of states in this band, but requires a more elaborate multitrace deformation than the solvable portion of the $T^2+\Lambda$ theory, which can be expressed step by step and treated as an algorithm generating the deformation \cite{Batra:2024kjl, Silverstein:2024xnr}.

Let's first review the implications of these choices when $\delta=0$. The matching point at $y=y_0$ is such that the square root vanishes for $\Delta_*=c/6$. In the CFT, for $\Delta \ge \Delta_*$, the density of states is given by the Cardy formula \cite{Cardy:1986ie}
\be\label{eq:cardy-density}
\rho_C(\Delta)=\exp \left(2 \pi  \sqrt{\frac{1}{3} c \left(\Delta-\frac{c}{12}\right)}\right)\,.
\ee
For $\Delta< \Delta_*$, the density of states is non-universal and much smaller than the Cardy result \cite{Hartman_2014}. This will play an important role below. Now, the role of $\delta$ is to produce an $O(1)$ window for the states around the Cardy level, whose density accounts for the microcanonical entropy; at large $c$ this was discussed in detail e.g.~in \cite{Mukhametzhanov:2019pzy}.
The gravitational dual of the $\Delta_*=c/6$ state is an AdS BH microstate at the Hawking-Page transition \cite{Hartman_2014, Witten:1998zw}, and the $T \bar T +\Lambda_2$ trajectory matches this state with a corresponding microstate of the de Sitter static patch, with the microstates subject to small rearrangements due to the matter portion of the full $T\bar T +\Lambda_2 + matter$ multitrace deformation.  We will denote the dressed and matter-adjusted density of states $\rho_{CO}$, and note that its main features (support over an O(1) band of $\simeq \exp(A/4G_N)$ energy levels) are not disturbed by the matter \cite{Batra:2024kjl, Silverstein:2024xnr}. 

In our 2-sided theory, the microscopic TFD state gives a reduced density matrix whose von Neumann entropy matches the microcanonical entropy. Using the above considerations for the spectrum of states, the one-sided partition function (or trace of the single-sided density matrix) is 
\be\label{eq:Zbeta1}
Z(\beta) = \int_{{\cal E}_{C,1}}^{{{\cal E}_{C,2}}} d {\cal E}\,\rho_{CO}\left({\cal E}\right) e^{-(\beta/L) {\cal E}}+ \int_{{\cal E}_0}^{{\cal E}_{C,1}}\,d {\cal E}\,\rho_{\rm non-univ}\left({\cal E}\right) e^{-(\beta/L) {\cal E}}\,.
\ee
The integral is over the deformed energies; ${\cal E}_{C,1}$ and ${\cal E}_{C,2}$ determine the band of deformed energies corresponding to $\Delta \in (c/6,c/6+\delta)$, and $\rho_{CO}({\cal E})$ is the dressed and matter-corrected Cardy density of states. The second term encodes the contributions of energy levels below the top band, whose density of states depends on details of the matter sector and is suppressed compared to $\rho_{CO}$.

Next, let us compare (\ref{eq:Zbeta1}) with the gravitational calculations above. The integrand $\rho_{CO} e^{-(\beta/L) {\cal E}}$ matches the exponential of the on-shell gravitational action with a Dirichlet wall, where ${\cal E}$ is proportional to the Brown-York energy (up to a local counterterm), and $\log \rho_{CO}$ is approximately the horizon area. In the constrained parametrization, this was given in (\ref{I-can-L0}).   Specifically, in terms of $L_0$, the entropy is the $L_0/4G$ term in \eqref{I-can-L0}, and the Brown-York energy is $\frac{1}{8\pi G_N \ell}\sqrt{L_0^2-L_c^2}$. 
Here we are applying the Brown-York energy $\int T^{00} \sim \int(K^{00}-h^{00} K)$ in the constrained path integral, so only partially off shell.  The standard derivation of the Brown-York stress tensor works on-shell, so the conservation equation suffers an off-shell correction proportional to the Einstein tensor:
\begin{equation}
    D_i T^{i0} \propto G^{0\mu}n_\mu=G^{0r}
\end{equation}
with $r$ our radial direction normal to the boundary, and $G$ the Einstein tensor.  This component is one of the constraints, which vanishes in our constrained gravitational path integral configurations even as we vary $L_0$ off shell.
{ Thus the partially off-shell stress tensor is still conserved.} Note that for the class of metrics (\ref{U1-sym-metric}) used for the constrained path integral, this notion of BY energy agrees with the variation of the constrained action with respect to the time component of the boundary metric. 

The microscopic expression for the partition function now explains how the UV completion resolves the instabilities noted in the analysis of (\ref{I-can-L0}). First, the unboundedness as $L_0 \to \infty$ is removed because from the point of view of the quantum model it would correspond to taking arbitrarily large undeformed energies above the matter-deformed Cardy level, which are absent from the spectrum.\footnote{In the gravitational side, this corresponds to placing a negative mass on the other side of the wall. This is allowed by the gravitational boundary conditions, but it is a self-consistent design choice of our Hamiltonian theory to truncate the spectrum to discard states with energies that ever go complex in our deformation.} This places an upper bound $L_0 \lesssim 2\pi \ell$. On the other hand, once $L_0$ is smaller than $2\pi \ell$, the microscopic dual implies that the density of states (what we called $\rho_{\rm non-univ}$ before) suppresses these contributions compared to the O(1) matter-deformed Cardy window $L_0 \sim 2\pi\ell$ captured by the first term in (\ref{eq:Zbeta1}) at temperatures above the HP transition.
These effects are respectively demonstrated by the dashed and dotted segments of the plot in Fig. \ref{fig-canonical-action}.

We also stress that in terms of (\ref{eq:Zbeta1}), which goes beyond the gravitational saddle point approximation, the canonical ensemble above the HP transition is explicitly equivalent to the microcanonical ensemble calculation inside the Cardy band. This is based on the upper bound on energies, and the fact that the density of states below the Cardy level (which includes the sparse light spectrum) is suppressed compared to (\ref{eq:cardy-density}).  In Sec.~\ref{sec-microcanonical} below, we will see that the microcanonical thermofield double does not suffer from a classical instability in the constrained path integral.

\subsubsection{Specific heat of de Sitter}

Similar considerations affect other thermodynamic observables. A known issue in de Sitter gravitational thermodynamics has been the fact that Schwarzschild de Sitter (SdS) black holes can dominate in the canonical partition function, but nevertheless have negative specific heat $C=\beta^2 \partial^2 \log Z/\partial \beta^2$ at the gravitational classical solution. See \cite{Draper:2022ofa} for a recent analysis of this point, and \cite{Draper:2022xzl, Banks:2024cqo} for proposed resolutions of this problem in terms of constrained states and a modified replica calculation. In what follows we will show how the properties of the microscopic spectrum explicitly avoid $C<0$.

Again for simplicity we illustrate this in $d+1=3$, where the Euclidean SdS$_3$ metric reads
\be
ds^2= \frac{r_h^2-r^2}{\ell^2}d\tau^2+ \frac{\ell^2 dr^2}{r_h^2-r^2}+r^2 d\phi^2\,.
\ee
This is not really a black hole, but rather the backreacted metric for a pair of particles at antipodal points (see e.g.~\cite{Spradlin:2001pw}).  Nevertheless, it is enough for our goal, since $r_h$ provides an independent parameter that can be used to change the temperature. The cosmic horizon patch now extends between $R_c<r<r_h$, and regularity at the horizon implies that $\tau$ has periodicity 
\be
\beta_h= \frac{2\pi \ell^2}{r_h}\,.
\ee
Therefore, the temperature at the wall $r=R_c$ is
\be
\beta_c= 2\pi \ell \sqrt{1-R_c^2/r_h^2}\,.
\ee
Rewriting the Brown-York energy as a function of the inverse temperature $\beta_c$ at the wall gives\footnote{Here we choose a counterterm that coincides with the choice in the dual theory.}
\be \label{eq-EBY-beta}
E_{BY}= \frac{2 \pi  R_c }{8 \pi  G_N \ell}\left(1+\sqrt{-1+\frac{r_h^2}{R_c^2}}\right)=\frac{R_c}{4 G_N \ell} \left( 1+\beta_c \sqrt{\frac{1}{4 \pi ^2 \ell^2-\beta_c^2}}\,\right)\,.
\ee
This implies a negative specific heat,
\be\label{eq:C1}
C= -\beta_c^2 \frac{\partial E_{BY}}{\partial \beta_c}=-\frac{\pi^2 \ell R_c \beta_c^2}{G_N}\left(\frac{1}{4 \pi ^2 \ell^2-\beta_c^2}\right)^{3/2}\,.
\ee
This cannot arise from a well-defined partition function $Z(\beta)$ with a positive density of states, which would always give
\be
C= \beta^2 \langle (E- \langle E \rangle )^2 \rangle  \ge 0\,.
\ee

In order to understand the origin of the $C<0$ inconsistency, it is useful to write it in terms of fluctuations of the entropy as a function of energy. Writing the
gravitational entropy $S=(2\pi r_h)/4G$ in terms of the Brown-York energy at the wall gives an expression for the microcanonical entropy 
\be\label{eq:Sgrav}
S_{\rm grav}({\cal E})= \sqrt{\frac{c}{3}} \sqrt{\frac{2}{y}-2 \pi  {\cal E}+\pi ^2 {\cal E}^2 y}\,.
\ee
Here $y=8G\ell/L_c^2$,  $c=3\ell/2G$ and $\mathcal{E}=E_{BY} L_c$, with $L_c=2\pi R_c$.
Then we can check that the above expression for the specific heat computed from the gravitational saddle coincides with
\be\label{eq:C2}
C= - \frac{(\partial S_{\rm grav}/\partial {\cal E})^2}{\partial^2 S_{\rm grav} /\partial {\cal E}^2}\,.
\ee
On the other hand, in terms of our microscopic expression (\ref{eq:Zbeta1}), we have
\be\label{eq:C3}
Z(\beta) \langle ({\cal E}- \langle {\cal E}\rangle)^2 \rangle=\int_{{\cal E}_{C,1}}^{{{\cal E}_{C,2}}} d {\cal E}\,\rho_{CO}\left({\cal E}\right)\,({\cal E}- \langle {\cal E}\rangle)^2 \,e^{-(\beta/L) {\cal E}}+ \int_{{\cal E}_0}^{{\cal E}_{C,1}}\,d {\cal E}\,\rho_{\rm non-univ}\left({\cal E}\right) \,({\cal E}- \langle {\cal E}\rangle)^2\,e^{-(\beta/L) {\cal E}}\,. 
\ee
We can now understand the origin of (\ref{eq:C2}): it arises from the gravitational analog of the first term in (\ref{eq:C3}), expanding $e^{S_{\rm grav}-\beta {\cal E}/L}$ to quadratic order, and then computing the energy fluctuation by saddle point in terms of the quadratic coefficient. 
However, this is valid if the energy band over which this term is valid is sufficiently large, such that the curvature of $S_{\rm grav}-\beta {\cal E}/L$ can be detected. And this is not the case. Instead, for $\delta \sim 1$ fixed at large $c$, the integral of the first term gives
\be
\frac{1}{Z(\beta)}\int_{{\cal E}_{C,1}}^{{{\cal E}_{C,2}}} d {\cal E}\,\rho_{CO}\left({\cal E}\right)\,({\cal E}- \langle {\cal E}\rangle)^2 \,e^{-(\beta/L) {\cal E}} \sim \delta^2 >0\,.
\ee
 Of course, since this term now gives an $O(1)$ contribution to $C$, the non-universal part of the spectrum may also contribute at this order. But this shows that the resolution of the specific heat negativity has a similar origin as the resolution of the $L_0$ instability.

To recap: we addressed the apparent instability of the saddle solutions, i.e.~the dS geometry filling in the bulk starting from the canonical TFD boundary (and similarly for the associated calculation of the one-sided thermal partition function).

We exhibited UV sensitivity of the canonical TFD in the relevant regime above the Hawking-Page transition, where the boundary size is of order the dS radius.  From the gravity-side perspective, UV issues arise from singularities if the RT surface goes off-shell and from QFT effects.
There is an interesting landscape of configurations that could dominate in the gravitational + matter path integral, somewhat analogous to the string/M theory landscape itself which depends on stringy and/or quantum physics beyond classical GR + matter \cite{Flauger:2022hie,Silverstein:2004id, Silverstein:2016ggb}.   

In this section, working from the boundary side, we showed that the instability is spurious, as the integration range is truncated on both sides of the saddle in accord with the known top heavy structure of our finite dressed energy spectrum. We also demonstrated that the same treatment renders the specific heat positive.

We turn now to the microcanonical version of the TFD, in a regime where the boundary size is small and the gravity-side UV sensitivity is less significant.   

\subsection{Microcanonical version:  stable saddle}\label{sec-microcanonical}

For the microcanonical version \cite{Marolf:2018ldl}, we define 
\begin{equation}\label{eq-mctfd}
    |\Psi_{\rm MCTFD}\rangle = {\cal N}\sum_i e^{-\beta_c E_i/2}f(E_i-E_{max})|E_i\rangle |E_i\rangle 
\end{equation}
where $f$ is well peaked at $E=E_{max}$.   This is again close to maximally mixed. We will take the window function $f$ to have a support contained in the $\delta$-band around the top level as defined above in \S\ref{subsec:UVcomplete}.

In this ensemble, there is no upper bound on the value of $\beta_c$ and hence $L_c$, in contrast to the bound (\ref{eq:HPbound}) we had for the canonical ensemble in order to be above the Hawking-Page transition.  Since the boundary can be small, the significant effect of the boundary stress energy on the region near the spatially maximal surface in dS \eqref{eq-UV-sensitivity-estimate} no longer arises. 

Interestingly, in this case we find that the microcanonical condition stabilizes the area of the central slice of the geometry at the classical level. This has been discussed in the higher dimensional context in \cite{Banihashemi:2022jys}.

The density of states with energy $E$ may be represented by a microcanonical path integral \cite{Brown:1992bq}:  
\be
\nu(E)=\int \mc{D}g_{\mu\nu}\, e^{-I_{\rm mic}}\, ,
\ee
where the functional integral is over metrics satisfying a fixed Brown-York energy (instead of temperature). 
The Brown-York energy \eqref{eq-BY-tensor}
can be written in terms of the trace of the extrinsic curvature of the spatial boundary as embedded in a constant-$\tau$ slice as 
\be
E_{BY}=-\frac{1}{8\pi G_N}\int \frac{d\phi}{2\pi} L(r_c) k \, , \quad \quad \text{where} \quad k=\frac{L'}{L}\, .
\ee
The microcanonical action can be derived by requiring that its variation yields the field equation once the system boundary size and the BY energy density ($\sim k=L'/L$) are held fixed, which amounts to fixing $L$ and $L'$ at the boundary. It can then be readily shown that 
$I_{\rm mic}$ 
differs from the canonical action $I$ by the last term in \eqref{I-bL}. 
Thus we get 
\be\label{eq:Imic}
I_{\rm mic} = -\frac{L_0}{4G_N}\, .
\ee
Recall that $L(r)=L_0 \cos r/\ell$ upon imposing the constraint.
Thus $\cos r_c/\ell=L_c/L_0$, and the microcanonical boundary condition becomes
\be\label{eq:Ebymicro}
E_{BY}=-\frac{1}{8\pi G_N}L'(r_c)=\frac{1}{8\pi G_N \ell}\sqrt{L_0^2-L_c^2} \equiv E\, .
\ee
The value $L_0$ is thus fixed once the BY energy is specified, so in this case we get a single contribution from the on-shell configuration. As before, let's denote this on-shell value by $L_h$, which is a function of $(E,L_c)$.

Similar to the path integral representation of the density of states, one can construct the path integral computing the norm of the state $|\Psi_{\rm MCTFD}\rangle$. As demonstrated in \cite{Marolf:2018ldl}, this is employed to compute the entanglement entropy, whose leading order term is indeed given by the horizon area in Planck units. By writing $L_0$ in terms of $E$ via (\ref{eq:Ebymicro}) and substituting into (\ref{eq:Imic}), we obtain the microcanonical entropy as a function of energy, $S_{\textrm{micro}}(E)= \log \nu(E)$. However, we should stress that with the window function $f$ above constrained to have support inside the Cardy band (of width $\delta$ for undeformed energies), the microcanonical result can basically only access $S_{\textrm{micro}}(\langle E \rangle)$, with $\langle E \rangle \sim E_{max}$. It does not have fine-grained information to detect energy variations within the band.
The apparent canonical instability of the cosmic horizon solution (before we bring in the UV effects discussed above) is somewhat analogous to small AdS black holes in higher dimensions, where realizing the saddle in microcanonical ensemble comes to rescue.

\subsection{A novel ensemble with a stable saddle}
\label{sec-new-bc}

As discussed in \S\ref{sec-UV-QFT-illustration}, one-loop quantum effects add terms to the effective action that can be incompatible with Dirichlet boundary conditions for the metric within the saddle-point approximation. In  the example of the scalar field we may find scalar boundary conditions such that the heat kernel coefficients $a_0, a_1, a_2$ in \eqref{eq:HKcoe3D} renormalize $G$ and $\Lambda$ in the classical theory, without disturbing the relative coefficients of the Einstein-Hilbert action and the Gibbons-Hawking-York term (see e.g., \cite{Jacobson:2013yqa}). So restricting to those terms still allows the Dirichlet boundary condition for the metric to yield a well-defined variational problem for the renormalized theory, with on-shell solutions. However, adding the next term, $a_3$ in \eqref{eq:HKcoe3D}, requires a new set of metric boundary conditions.
Varying the total action including the one-loop corrections for the class of geometries \eqref{U1-sym-metric}, we get 
\be
\delta (I_{\rm cl}+I_{\rm QFT})= \text{terms giving EOM} -\frac{1}{4G_N} (b' \delta L +L' \delta b)_c + \log (\epsilon)\, \delta  a_3|_c \, ,
\ee
where $a_3$ is proportional to the integral given in \eqref{eq:HKcoe3D}, and the subscript $c$ refers to the values at the cutoff. In what follows we denote the last term above as $\int d^2x \, \delta Q$, and keep in mind that the couplings $G$ and $\ell$ are the renormalized ones. 

One way to incorporate the quantum effects into  consistent boundary conditions for the metric is to make the term $(b' \delta L +L' \delta b)$ a total differential. 
We notice that this term written covariantly is indeed $\int \! d^2x \, \Pi^{ab} \delta h_{ab}$, where $\Pi^{ab}$ is the conjugate momentum to the boundary induced metric $h_{ab}$, defined from the extrinsic curvature tensor $K^{ab}$ by
\be
\Pi^{ab}=\frac{1}{16\pi G_N}\sqrt{h} \left ( K^{ab} - K h^{ab} \right  )\, .
\ee
The variation of the action is thus
\be \label{eq-DBC-variation}
\delta (I_{\rm cl}+I_{\rm QFT})= \text{EOM} +\int \! d^2x \, (\Pi^{ab} \delta h_{ab} + \delta Q ) = \text{EOM} +\int \! d^2x \, \left ( \delta \Pi - h_{ab} \delta \Pi^{ab} +\delta Q \right )\, ,
\ee
where $\Pi=h_{ab}\Pi^{ab}$ is the trace of the conjugate momentum. It is then clear that fixing $\Pi^{ab}$ and $\Pi+Q$ extremizes the total action at a solution to field equations.

This observation suggests defining a new ensemble even at the classical level, namely by fixing $\Pi^{ab}$ and $\Pi$. 
For the class of geometries \eqref{U1-sym-metric} the non-zero  momentum components and its trace are
\be
\Pi^{\tau \tau} =-\frac{1}{16\pi G_N}\frac{L'}{b} \quad , \quad \Pi^{\phi \phi} =-\frac{1}{16\pi G_N}\frac{b'}{L} \quad , \quad \Pi = -\frac{1}{16\pi G_N} (bL'+Lb')\, .
\ee
We fix the momentum in a specific way that is parameterized by a single constant $A$, so that we have 
\be\label{eq-Qcorrected-bc}
L'_c=-\frac{A}{\ell} b_c \quad , \quad b'_c=\frac{1}{A \ell} L_c \, .
\ee
Then fixing the trace amounts to 
\be \label{eq-trace-fixing}
-\frac{1}{8G_N \ell}\left (\frac{1}{A }L_c^2-A b_c^2 \right )= B\, .
\ee
Here $A$ is a dimensionless positive\footnote{It is chosen to be positive to accommodate cosmic-type solutions, in which the size of the transverse space decreases as the boundary is approached from the bulk.} constant, and together with $B$ they give the fixed boundary data. Note that the momentum components are not fixed independently, so there in no over-determination. It can be readily checked that the above conditions uniquely specify the solutions to the full set of Einstein equations.\footnote{Namely, $L(r)=L_0 \cos r/\ell$ and $b(r)=\ell \sin r/\ell$ with $L_0/\ell=A$ and $r_c$ given by $\cos 2r_c/\ell = -8GB/A\ell$.}
Using the solution to the constraint $L=L_0 \cos r/\ell$ and expressing $b$ in terms of $L'$ the action becomes
\be \label{eq-new-bc-action}
I=-\frac{L_0}{4G_N}-\frac{1}{4G_N}bL'|_c=-\frac{L_0}{4G_N}+\frac{1}{8G_N \ell \, A} L_0^2+ B \, ,
\ee
which has a local minimum at $L_0/\ell = A$. 

This boundary condition is easily generalizable to the case of quantum-corrected action, where instead we have $\Pi+Q = \,$const., and the final action is again in the form \eqref{eq-new-bc-action}.
Another remark is that we could define the classical theory with the Einstein-Hilbert action with no additional boundary terms, and obtain the same result classically. To see it more explicitly, we note that the variation of the (Euclidean) Einstein-Hilbert action in $d+1$-dimensional bulk is
\be
\delta I_{\rm EH}= \text{EOM}+\frac{d-3}{d-1}\int_{\partial \mathcal{M} } \delta \Pi - \int_{\partial \mathcal{M} } h_{ab} \delta \Pi^{ab} \, ,
\ee
thus we may start from $I_{\rm EH}$ and impose the above boundary conditions to define the ensemble. It is clear that any additional term proportional to $\int_{\partial \mathcal{M} } \! \Pi$ (e.g., the GHY term) does not change the $L_0$ dependence of the action; it only shifts the action value by a constant.

This modification of the gravitational boundary conditions inspired by quantum matter effects has interesting properties.  For example from \eqref{eq-Qcorrected-bc} it relates energy to {\it inverse} temperature, as well as relating pressure to volume.
This is a novel ensemble which we expect does not change the count of states (entropy): we can formulate the TFD corresponding to these boundary conditions with window functions as in \cite{Marolf:2018ldl} without affecting the count of states in the top band of levels.

\section{Incorporating additional energy levels -- with matter generating tall geometries -- from a constrained Hamiltonian system}\label{sec-constraint-onto-joined-states}

In the previous section, we derived a correspondence between a highly entangled state and the connected geometry of empty de Sitter with two timelike boundaries.
In this section, we will outline the construction of a finite quantum system that incorporates connected geometries with matter.  This theory will contain energy bands dual to such semiclassical geometries.
A key feature of the $\Lambda_{d+1}>0$ theory is the fact that positive-energy matter induces tall geometries, a fact that extends to our case of interest with timelike boundaries as we saw above in \S\ref{sec:tall-example-geometries}.  

The Hamiltonian of this system may generally be written as
\begin{equation}\label{eq:H1}
    H = \sum_n E_n |E_n\rangle \langle E_n|\, .
\end{equation}
In gravity language, the expectation value of the energy is given by the semiclassical expectation value of the Brown-York energy
\begin{equation}
  \langle H \rangle \simeq \langle E_{BY}\rangle_{GR+EFT} =\sum_{\rm boundary~components} E_{BY, \text{component}}\, ,
\end{equation}
up to corrections controlled by quantum gravity effects going beyond GR+EFT, which will be defined by our theory in the same sense as laid out in the one-sided case in \cite{Batra:2024kjl, Silverstein:2024xnr}.    

We start from two copies of our $T^2 + OO + \textit{uplift} + \Lambda_d$ dressed system \cite{Coleman:2021nor, Batra:2024kjl, Silverstein:2024xnr}.  In this paper our main interest is in smooth spacetimes, in general with matter, extending beyond a given bounded observer patch. We will construct a theory which includes those and does not need to include arbitrary product states within the doubled Hilbert space
\be\label{eq-product-hilbert}
\mc H_{\rm product} = \mc H_L \otimes \mc H_R\,.
\ee
 This will define a map from the product Hilbert space to the physical or joined Hilbert space
\be
{\mathcal C}: \mc H_{\textrm{product}} \to {\mc H}_{\textrm{phys}}
\ee
Individual microstates of our system are generically singular at or past the horizon, similarly to black hole microstates (of which de Sitter microstates are dressed versions \cite{Coleman:2021nor, Silverstein:2024xnr}).      
Focusing instead on smooth extensions of the spacetime, we can define our spectrum to only include them.  In order to relate this to a description in terms of our original pair of dressed Hamiltonians, this will lead us to introduce some constraints on the Hilbert space which will naturally encode the operator redundancies noted for tall geometries in the introduction.  We reiterate that smoothness also fits with a finite entropy, since generic EFT excitations introduced asymptotically at early or late times (including those in the metastable de Sitter landscape) lead to singular behavior in the intermediate evolution \cite{Giddings:2007nu}.  

\begin{figure}[h]
    \centering
    \includegraphics[width=0.7\linewidth]{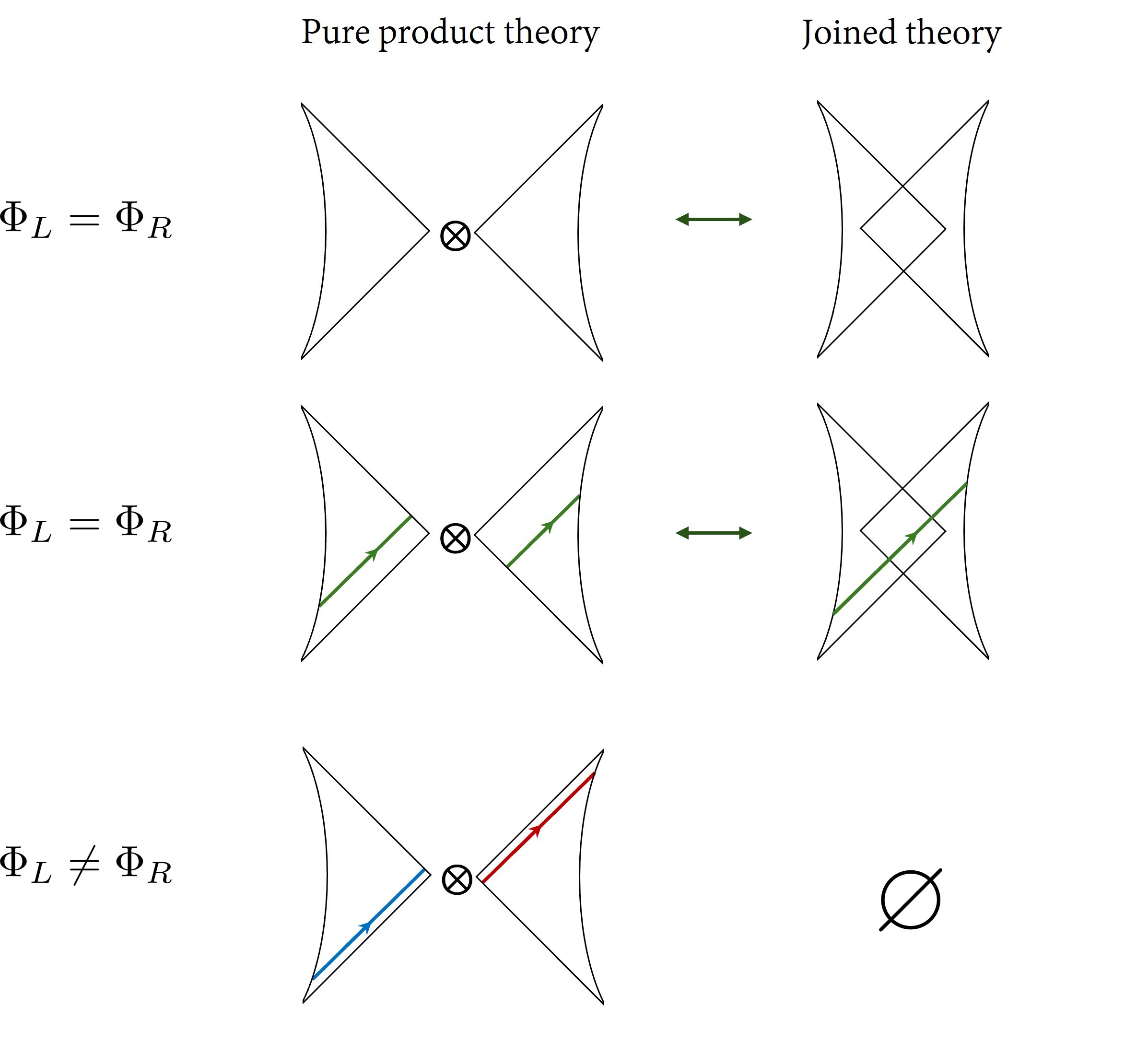}
    \caption{The product of two of the dressed quantum systems derived in \cite{Batra:2024kjl, Silverstein:2024xnr}  (left panel) includes energy bands that have no counterpart in the spectrum of our joined theory, as in the last row.  To define a theory that restricts to consistently joined energy bands, like in the first and second rows, we project them out with the constraint ${\cal C}$.  This automatically incorporates the operator redundancy between $\phi_L$ and $\phi_R$ on the common diamond of the tall joined theory:  it forbids independent application of L and R bulk operators. Here and in subsequent figures, the arrows depict matter excitations.}
    \label{fig:constrainedtheory}
\end{figure}

Consider the product spectrum in the right panel of figure \ref{fig:constrainedtheory}.  
It contains energy levels arising from completely independent L and R wedges.
For example, it would include product states consisting of tall geometries whose matter excitations do not agree on what would be the overlap region of a joined tall geometry.  For the purpose of our theory of extended geometries, we can project these out with a constraint ${\cal C}$:
\begin{equation}
    {\cal C} |\text{inconsistently-joined} \rangle =0.
\end{equation}
Operationally, the full Hamiltonian is given by
\be
H= {\mc C} (H_L+H_R) {\mc C}
\ee
and we choose ${\mc C}^2=\mc C$.
This satisfies two key properties:

1)  It commutes with the Hamiltonian for the product theory
\begin{equation}
    [H_L + H_R, {\cal C}]=0;
\end{equation}
so the projection onto the physical Hilbert space is preserved under time evolution.

 2) It encodes the requisite operator redundancies, eliminating independent action of $\phi_L$ and $\phi_R$ in the overlapping region of the joined diamonds:  
\be\label{eq-op-constraints}
\phi_L = \phi_R ~~~~~ \text{on  overlap  diamond}
\ee
We can see this from the fact that independent application of $\phi_L$ separately from $\phi_R$ on the top band of energy levels would produce inconsistently joined states, which are projected out.  So these operators do not exist independently; they are redundant. 
The fact that there are these redundancies means that overall there is not a clean separation between the two theories; their Hilbert space does not factorize.  However, we note that this  does not affect the $\simeq A/4G_N$ worth of entanglement entropy in the top band of energies discussed above in \S\ref{sec-top-joining-TFD-etc}.\footnote{{We thank G. Penington for a discussion of this.}} 

\subsection{States where L(R) causal wedge overlaps with the R(L) boundary}

The full spectrum of joined geometries includes ones where the opposite boundary is in the causal wedge of a given boundary.  The energy remains the sum of the two Brown-York energies.  The corresponding block of the Hamiltonian then can be written as
\begin{equation}
    H = H_L + \int_{S^{d-1}}T_{00, BY}^R[T_{BY, L}]
\end{equation}
where the second term contains the R Brown-York stress tensor at the position of the R boundary as reconstructed in terms of the L theory's stress-energy tensor, integrated over the boundary's spatial sphere.  Within the causal wedge of the L boundary, $T^R_{00, BY}$ is a bulk operator, the extrinsic curvature, reconstructed in terms of L boundary operators with an appropriate kernel \cite{Hamilton:2006az}.

\subsection{Nonzero commutators in tall states}
\label{sec:nonzero_comms}

Let us consider computing the quantity 
\begin{align}
\label{eq:comm}
    \ev{\psi_t|[O_L(t),O_R(0)]|\psi_t}
\end{align}
from the boundary theory point of view, where $\psi_t$ represents a tall state. Without imposing any constraints on the Hilbert space, since the left and right systems are decoupled, this quantity is zero.   

In the constrained system, we have operators ${\cal C} O {\cal C}$ acting within the physical Hilbert space.  To see that $\bra{\psi_t}[\mathcal{C}O_L(t_L)\mathcal{C},\mathcal{C}O_R(0)\mathcal{C}]\ket{\psi_t}$ can be nonzero, we start by writing 
    \begin{equation*}
        \ev{\psi_t|[\mathcal{C}O_L\mathcal{C},\mathcal{C}O_R\mathcal{C}]|\psi_t}=\ev{\psi_t|O_L\mathcal{C}O_R-O_R\mathcal{C}O_L|\psi_t}
    \end{equation*}
This will still vanish if either of $O_L$ or $O_R$ commutes with the projector ${\cal C}$.  We would like to understand for which operators they both fail to commute with ${\cal C}$ so that the commutator need not vanish.

\begin{figure}[h]
    \centering
    \includegraphics[width=0.5\linewidth]{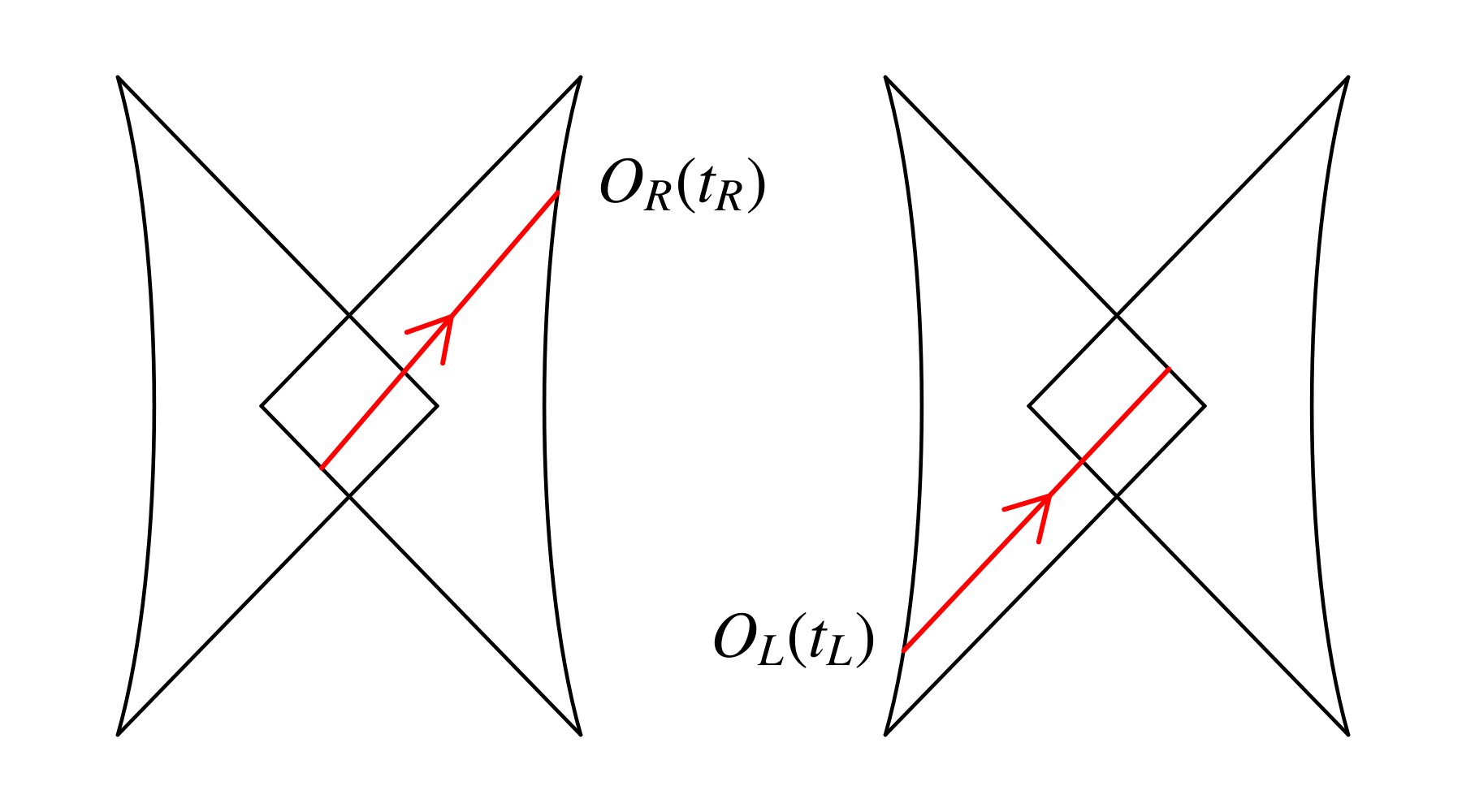}
    \caption{If we act with $O_R(t_R)$ for sufficiently late $t_R$ on the bra, or with $O_L(t_L)$ with $t_L$ at a sufficiently early on the ket, this corresponds via \eqref{eq-operator-growth-general} to an action at $t=0$ that changes the configuration in the overlap wedge. This fact is essential for showing that the quantity $\ev{\psi_t|[O_L(t),O_R(0)]|\psi_t}$ is non-zero in a tall state, see \S \ref{sec:nonzero_comms}.}
    \label{fig:comm_overlap}
\end{figure}

Working in the Heisenberg picture, let us consider $O_L(t_L, \Omega_L)$ and $O_R(t_R, \Omega_R)$ to be simple local operators at early and late times respectively, along the L and R boundaries, where $\Omega$ denotes the coordinates on the spatial sphere of dimension $d-1$ on the boundary.  Expressed in terms of operators at time $t=0$ (the moment of time reflection symmetry of the doubly bounded geometry), given sufficient time evolution, this is a complex operator
\begin{equation}\label{eq-operator-growth-general}
    O_R(t_R) = O_R(0) + i t_R [H, O_R(0)]-\frac{t_R^2}{2}[H, [H, O_R(0)] ]+\dots 
\end{equation}
and similarly for $t_L$.  For sufficiently late $t_R$, acting with this operator will affect the overlap diamond of the tall system that is centered at $t=0$, and similarly for sufficiently early $t_L$ (see figure \ref{fig:comm_overlap}). Acting with $O_L$ or $O_R$ independently within the would-be overlap diamond can ruin the consistent overlap, taking the system out of the physical Hilbert space, meaning $[O, {\cal C}] \ne 0$.  This makes possible a nonzero commutator.  
Specifically, the early operator acting on the ket $|\psi_t\rangle $ can create a particle propagating to the overlap diamond, and the late operator acting on the bra $\langle \psi_t|$ will absorb a particle.  Here we are using the fact that the local boundary operators are given by the radial field momentum, a weakly interacting bulk field.  
Conversely, at times $t_L, t_R$ near $t=0$, however, the operator will not be complex enough to impact the overlap diamond, and the commutator will be zero. 
The commutator will also be zero if both operators are early or late, since the one acting on a bra will absorb not create a particle and hence will commute with ${\cal C}$.  

In section \ref{sec-bulk-reconstruction} we turn to more details of the HKLL reconstruction procedure in our system, which ensures the correct commutator structure in the bulk.    
\subsection{Constrained Hilbert space in terms of Euclidean path integral with boundaries}

The TFD state can be represented gravitationally in terms of a Hartle-Hawking state constructed using Euclidean evolution. Without Dirichlet walls, this is a `no-boundary' state. The construction can be extended to produce `yes-boundary wavefunctions' corresponding to averages of microstates in the top energy band of the dual quantum description, as discussed in Sec. \ref{subsec:preparation}. Similar Euclidean path integrals could be used to construct the gravitational duals of averages of lighter microstates in the sparse part of the spectrum.

For illustrative purposes, let us consider a simple toy model for a tall dS geometry with metric\footnote{See e.g.\cite{Leblond:2002ns} for solutions with explicit bottom-up scalar field potentials.}
\be\label{eq:tall1}
ds^2= \frac{\ell^2}{\cos^2 T}(-dT^2+ \mu_0^2 (d\theta^2+ \sin^2 \theta \,d\Omega_{d-1}^2))\,.
\ee
For $\mu_0=1$ this is global $dS_{d+1}$ in conformal coordinates, with $-\pi/2\le T \le \pi/2$ and $0 \le \theta \le \pi$. On the other hand, when $\mu_0<1$, we obtain a `tall' geometry. 
We consider two Dirichlet walls at $\theta_c(T)$ and $\pi-\theta_c(T)$, with fixed proper size
\be
R_c= \ell \,\mu_0\, \frac{\sin \theta_c(T)}{\cos T}\,.
\ee

\begin{figure}[h]
    \centering    \includegraphics[width=0.7\linewidth]{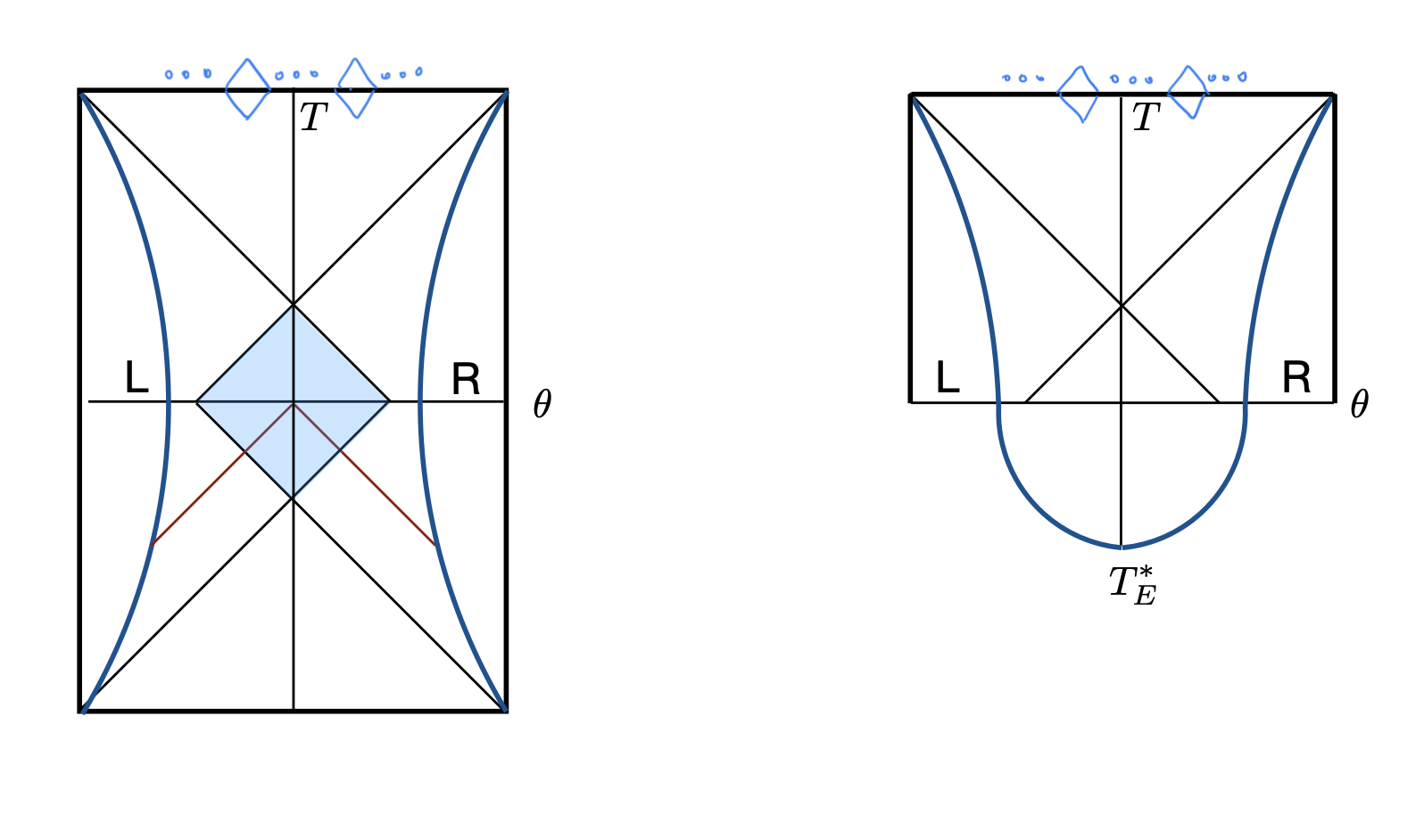}
    \caption{Left panel: tall dS geometry (\ref{eq:tall1}) with L and R boundaries of fixed proper size. The blue shaded causal diamond is at the intersection of the L and R causal wedges.  The red lines are null geodesics corresponding to the shortest past times at which emitted signals from the L and R boundaries can meet in the  time-reflection symmetric slice of the bulk. Right panel: state at $T=0$ is created by Euclidean evolution with background metric (\ref{eq:tall2}), and the L and R boundaries meet at $\theta=\pi/2$ and $T_E=T_E^*$. In pure dS, the Euclidean path integral gives the entangled HH state. In tall dS, we interpret it as creating a constrained state from the product of L and R Hilbert spaces. This corresponds to the tall version of Fig.\ref{fig:TFD_bdry}.}
    \label{fig:tall_eucl}
\end{figure}

We take now a bulk scalar field $\phi$ and create small excitations around the tall geometry. As shown in the left panel of Fig. \ref{fig:tall_eucl}, the existence of a bulk causal diamond causally connected to both boundaries puts some constraints on the type of allowed excitations, related to imposing appropriate regularity conditions in the diamond.

It is useful to create these excited states in terms of a Euclidean path integral,
\be
\Psi_{\rm excited}= \int D\phi\,e^{-S(\phi)-J \phi}\,,
\ee
where $J \phi$ is a schematic for past Euclidean insertions. The path integral is performed in a gravitational background that is the Euclidean continuation of (\ref{eq:tall1})
\be\label{eq:tall2}
ds_E^2= \frac{\ell^2}{\cosh^2 T_E}(dT_E^2+ \mu_0^2 (d\theta^2+ \sin^2 \theta \,d\Omega_{d-1}^2))\,,
\ee
and the range for the Euclidean time evolution is $-T_E^*\le T_E \le 0$.
The Euclidean time integral stops at $T_E=0$, where the real time evolution continues. Furthermore, the initial Euclidean time $-T_E^*$ arises from the two boundaries meeting at the equator $\theta=\pi/2$, namely
\be
\cosh T_E^*= \frac{\ell \mu_0}{R_c}\,.
\ee
This depends explicitly on the size of the Dirichlet wall and the tallness parameter $\ell \mu_0$. This is depicted in the right panel of Fig. \ref{fig:tall_eucl}.
In summary, while in pure dS the Euclidean path integral gives the entangled HH state, in tall dS we interpret it as creating a constrained state from the product of L and R Hilbert spaces. Although here we have considered small excitations on top of a given tall dS geometry (using a bulk matter field), in principle this gravitational construction of the constrained physical states can be performed more generally. As we discussed before, this will certainly require UV information from the dual theory and its spectrum. In this direction, it would be interesting to explore if a version of Cauchy slice holography \cite{Araujo-Regado:2022gvw, Soni:2024aop, shyam2026dscauchy} could be used for this purpose.

\section{Aspects of bulk reconstruction}
\label{sec-bulk-reconstruction}

In this section we comment on some aspects of bulk reconstruction in our joined system. We outline how it may be carried out using HKLL \cite{Hamilton:2005ju, Hamilton:2006az, Kabat:2011rz, Heemskerk:2012mn}, both in the empty \cite{Lewkowycz:2019xse} and tall de Sitter cases. We then outline some interesting nuances related to the resolution capacity of this reconstruction. The main point we want to emphasize here is that in tall de Sitter space, reconstructing the information in the causal wedges of the two boundaries is enough to generate the entire spacetime, including the future wedge.

Let us first consider reconstructing a bulk scalar $\phi(X)$ in the empty static patch with a boundary, where $X$ stands for $(r,t,\Omega)$. The analysis is similar to the AdS case, we essentially follow the discussion in \cite{Harlow:2018fse}. This field has a mode expansion given by
\begin{align}
\label{eq:gen_mode_exp}
    \phi(X)&=\sum_i(f_i(r,t,\Omega) a_i+f_i^*(r,t,\Omega) a_i^\dagger)\\
    &=\sum_{l\vec m}\int_0^\infty d\omega\ (f_{\omega l\vec m}(r,t,\Omega) a_{\omega l\vec m}+f^*_{\omega l\vec m}(r,t,\Omega) a^\dagger_{\omega l\vec m})
    \label{eq:mode_exp}
\end{align}
where $i$ ranges over the two independent solutions to the wave equation for $\phi(X)$. In going to the second line we used the symmetries of the static patch to write $f_{\omega l\vec m}=\psi_{\omega l}(r) e^{-i\omega t} Y_{l\vec m}(\Omega)$.
For $\psi_{\omega l}(r)$ we choose the linear combination of solutions to the wave equation that satisfies Dirichlet boundary conditions at the boundary. Since we are not imposing any additional boundary conditions at the horizon, $\omega$ is allowed to take continuous values.
Now, we can also expand the field $O(x)$ (where $x$ is the coordinate on the boundary) that corresponds to the boundary operator dual to $\phi$ (the radial momentum $\Pi\sim-\partial_r\phi$) in terms of creation and annihilation operators:
\begin{align}
\label{eq:bdry_mode_exp}
    O(x)&=\sum_i(F_i(t,\Omega) a_i+F_i^*(t,\Omega) a_i^\dagger)\\
    &= \sum_{l \vec{m}} \int_0^\infty d\omega \ N_{\omega l}(r_c) e^{-i\omega t} Y_{l \vec m}(\Omega) a_{\omega l\vec{m}}+\text{h.c.}.
\end{align}
Now, we may write
\begin{align}
    a_{\omega l\vec{m}}=\frac{1}{N_{\omega l}} \int dt\ e^{i\omega t} \int d\Omega \ Y_{l \vec{m}}^* (\Omega) O(x),
\end{align} 
and similarly for $a_{\omega l\vec{m}}^\dagger$.
Plugging this back into the expression for $\phi(X)$ in Eq. (\ref{eq:mode_exp}), we get the HKLL reconstruction formula in the static patch:
\begin{align}
\label{eq:HKLL_with_kernel}
    \phi(X)&=\int dt'd\Omega' K(r,t,\Omega;t',\Omega')O(x),\\
    K(r,t,\Omega;t',\Omega')&=\sum_{l\vec m}\int_0^\infty \! d\omega \left( N_{\omega l}^{-1} f_{\omega l\vec m}(r,t,\Omega) e^{i\omega t'}Y_{l\vec m}^*(\Omega')+ N_{\omega l}^{-1} f_{\omega l\vec m}^*(r,t,\Omega) e^{-i\omega t'}Y_{l\vec m}(\Omega')\right).
\end{align}
In deriving this formula we exchanged the order of the integration over the boundary and the integration/sum over modes. It is important that despite this maneuver, $\phi(X)$ still make sense as a distribution \cite{Bousso:2012mh,Morrison_2014}. 
Having reconstructed both static patches using the above procedure, we may then reconstruct the future wedge as follows. Since we have reconstructed the scalar field on the $t=0$ slice of the full geometry, and know the boundary conditions on the two boundaries, we may time evolve using the bulk equations of motion (which we assume knowledge of) to obtain the scalar field configuration in the future wedge. 

\begin{figure}
    \centering
    \includegraphics[width=0.7\linewidth]{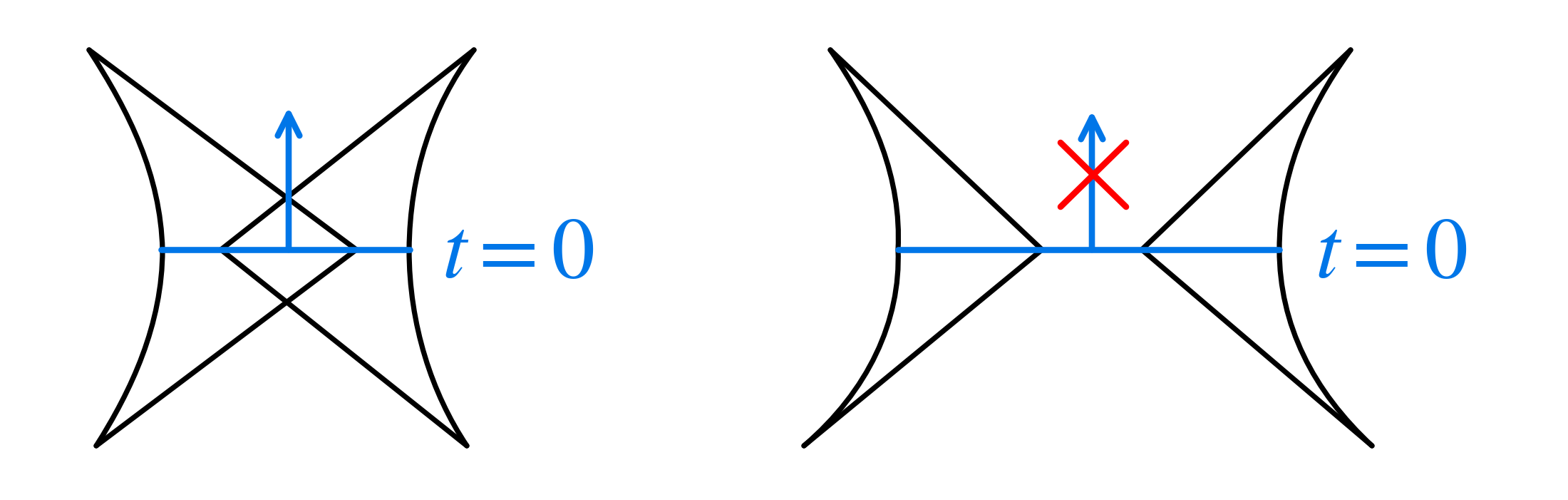}
    \caption{HKLL reconstruction is more powerful in de Sitter space with matter (left), since it allows us to reconstruct the full $t=0$ slice. Using the equations of motion, we can then reconstruct the future wedge. This is in contrast to anti de Sitter space with matter (right), where causal wedge reconstruction is not sufficient.}
    \label{fig:hkll}
\end{figure}

In the general case with matter in the static patch, Eq. (\ref{eq:gen_mode_exp}) still holds and we can expand the field in the bulk in terms of creation and annihilation modes. Similarly, we may also expand the field that corresponds to the dual operator as in Eq. (\ref{eq:bdry_mode_exp}). An HKLL reconstruction formula then exists if $F_i$ and $F_{i'}^*$ are orthogonal and it is possible to write \cite{Bousso:2012mh}
\begin{align}
    a_i=\int_{\partial\mathcal{M}} F_i^* O(x).
\end{align}
We want to emphasize two take-home messages here. The first is that compared to anti de Sitter space, HKLL reconstruction gives us more in de Sitter space (see figure \ref{fig:hkll}). This is because in the general situation with matter, HKLL enables us to reconstruct the entire $t=0$ slice in dS. This is in contrast with AdS where the causal wedges cover only part of the $t=0$ slice. 

\begin{figure}[h]
    \centering
    \includegraphics[width=0.5\linewidth]{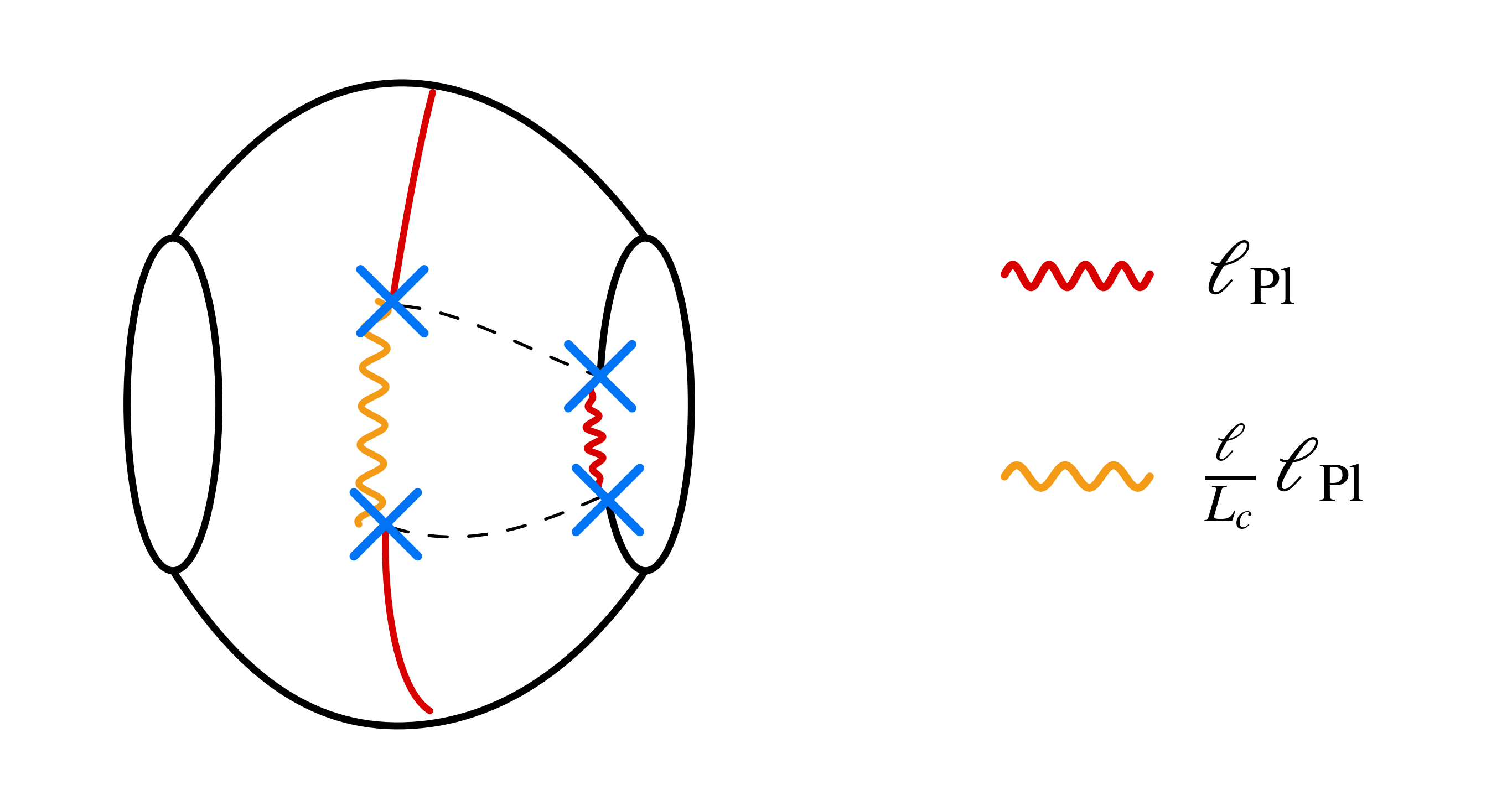}
    \caption{An illustration of the fact that reconstructing information probing scales of $\mathcal{O}(l_{\text{Pl}}\ell/L_c)$ in the bulk uses information at the Planck scale on the boundary. To reconstruct bulk information probing these scales, we consider correlation functions of operators separated by a distance of the order of this scale and located at $r\sim \ell$ (with the operators inserted at the blue crosses in the bulk). These are reconstructed in terms of correlation functions on the boundary using HKLL (with operators inserted at the blue crosses on the right boundary). The above picture is a spatial slice of dS with two boundaries,which are the ovals at the left and right sides. }
    \label{fig:res}
\end{figure}

The second is related to the resolution ability of HKLL reconstruction. Suppose we wish to compute the bulk two point function $\ev{\phi(r,t,\Omega_1)\phi(r,t,\Omega_2)}$ from the boundary two point function using 
\begin{align}
    \ev{\phi(r,t,\Omega_1)\phi(r,t,\Omega_2)}=\int dt'd\Omega' K(r,t,\Omega_1;t',\Omega') \int dt''d\Omega'' K(r,t,\Omega_2;t'',\Omega'') \ev{O(t',\Omega')O(t'',\Omega'')}.
\end{align}
For a given angular separation in the bulk $\Delta\Omega_{12}=\Omega_1-\Omega_2$, the above integral receives the largest contributions from the same separation on the boundary $\Omega'-\Omega''\sim \Delta\Omega_{12}$. This is because $K(r,t,\Omega_1;t',\Omega')$ is largest when $\Omega_1=\Omega'$, and $K(r,t,\Omega_2;t',\Omega'')$ is largest when $\Omega_2=\Omega''$ (a non-zero $\Omega_1-\Omega'$ or $\Omega_2-\Omega''$ increases the proper distance between the bulk and boundary points). So we can approximately write the above integral as
\begin{align}
    \ev{\phi(r,t,\Omega_1)\phi(r,t,\Omega_2)} \sim \int dt'\int dt'' K(r,t,\Omega_1;t',\Omega_1) K(r,t,\Omega_2;t'',\Omega_2) \ev{O(t',\Omega_1)O(t'',\Omega_2)}.
\end{align}
Additionally, the above integral is peaked when $t'=t''$, since that is when $O(t',\Omega_1)$ and $O(t'',\Omega_2)$ are closest to each other.
So, if we wish to resolve two points in the bulk that are separated by a proper distance $r \Delta \Omega_{12}$ on the sphere, we need to use boundary data involving points that are separated by a proper distance $L_c \Delta \Omega_{12}<r \Delta \Omega_{12}$.\footnote{This step should be understood as a resolution estimate rather than an
exact replacement of the HKLL integral.  It uses the fact that, for the
modes that dominate short-distance angular resolution, the smearing
kernel is peaked near coincident angular positions and equal boundary
times.  The omitted terms are the tails of the smearing functions and
low-angular-momentum contributions.  These terms are important for the
exact correlator,
but they do not change our parametric conclusion regarding angular resolution.} If $r$ is near the de Sitter horizon so that $r \sim \ell$, then the information probing scales that are parametrically larger than $l_{\text{Pl}}$ (by a factor of $\ell/L_c$) in the bulk depends on information at the Planck scale on the boundary. See figure \ref{fig:res} for an illustration of this point. However, since the metric is frozen on the boundary owing to the Dirichlet boundary conditions, this may not be a strong constraint, but may instead be an illustration of how we can fill in quantum gravity effects in the bulk using the information in the finite quantum system on the boundary.

As we discussed above in \S\ref{sec-constraint-onto-joined-states}, an interesting aspect of HKLL reconstruction in this theory is for the case of states with matter. This is the redundancy of reconstruction of local operators in the overlap diamond - their reconstructions on the left or right boundaries must be equivalent, imposing a non-trivial relation between operators of the left and right theories.

\section{Discussion}

Altogether, we have developed a holographic formulation of spacetime including cosmological future wedge in terms of two timelike boundaries excising diametrically opposite poles of de Sitter. This finite Hamiltonian system is built as a constrained combination of two copies of the one-sided theories developed in \cite{Coleman:2021nor, Batra:2024kjl, Silverstein:2024xnr}.   

The gravitational backgrounds incorporating the future wedge are generated by two types of energy levels in our theory:  (i) a thermofield double state built by near-maximal entanglement of the top band of microstates in the one-sided theories \cite{Coleman:2021nor, Batra:2024kjl, Silverstein:2024xnr}, and (ii) lower-energy tall states, obtained via a constraint imposed on the doubled Hilbert space.

To match the basic thermodynamics  -- in particular the entanglement entropy between the two sectors -- has required addressing within our framework the fact that the de Sitter horizon is a maximum spatially and that the de Sitter solution is a maximum in the pure gravity path integrand.  A priori, these properties appear to threaten the dominance of the saddle for the solution and the RT surface \cite{Shaghoulian:2021cef} in comparison to off-shell contributions, despite the fact that its prediction for the basic thermodynamic quantities (energy and entropy) are correct.    
We resolved this tension in several ways, applicable to different cases.  
The constrained path integral for the microcanonical TFD has a stable integrand, yielding $A/4G_N$ as the entropy.  The canonical TFD in the regime (above Hawking-Page) where the joined geometry is the dominant saddle suffers significant UV sensitive corrections; we showed explicitly how those can strongly affect the question of stability of the saddle for the bounded system.  In the end, the strongly top-heavy structure of our dressed energy spectrum, combined with the relation between the unstable direction and the energy, removes the would-be stronger off-shell contributions.

Our work is motivated in part by the genericity of timelike boundaries.   There is no known theoretical constraint on the spatial topology of our universe, rendering it potentially ill-advised to focus entirely on one very special case such as the closed universe. (A similar argument supports the study of more general topological features, both in $4d$ and in the extra dimensions of string theory).\footnote{The focus on very special topologies like closed universes (and internal Ricci flat or positive-curvature spaces) -- seems to be an example of model collapse, or perhaps a mathematical form of main character syndrome. }  It is equally important to note that their presence simplifies the problem of cosmological holography in several ways, particularly in the absence of dual gravity and the associated existence of multiple energy levels.  A trade-off in complication are the subtleties and UV sensitivities of the timelike boundaries themselves, which however is a subject that is progressing strongly \cite{Anninos:2011zn, Anderson:2006lqb, An:2021fcq, Anderson:2010ph, Liu:2024ymn,Liu:2025xij, Andrade:2015gja, Anninos:2022ujl, Anninos:2023epi, Banihashemi:2022htw, Banihashemi:2022jys, Banihashemi:2024yye,Banihashemi:2025qqi, Anninos:2024wpy, An:2025rlw, An:2025gvr, An:2025cbs,  Parvizi:2025wsg,Ahmadain:2024hgd,Ahmadain:2024uyo,Galante:2025emz, Galante:2025tnt,Anninos:2025zgr}. 

A consequence is that there are multiple possible states in cosmology.  There remains a rich interplay of theory and experiment at an appropriately model-dependent level \cite{Flauger:2022hie, Silverstein:2016ggb} (including the study of this sort of topology phenomenologically as initiated in  \cite{Philcox:2025faf}).

\noindent{\bf Acknowledgements.}
We thank Ayngaran Thavanesan, Vasu Shyam, and Ronak Soni for useful discussions in this area including another method \cite{shyam2026dscauchy} to access the future wedge. 
We are also grateful for useful discussions or correspondence with
Michael Anderson, Dio Anninos, Yiming Chen, Xi Dong, Ted Jacobson, Juan Maldacena, Don Marolf, Geoff Penington, XiaoLiang Qi, Pratik Rath, Harvey Reall, Jorge Santos, Edgar Shaghoulian, Douglas Stanford, and Aron Wall, along with many participants of the December 2025 Simons Center workshop ``Timelike Boundaries in Classical and Quantum Gravity", the September 2025 workshop ``Modern Trends in Gravity and Black Holes" at the University of Crete, and the December 2025 IAS workshop on ``Quantum Aspects of Black Holes and Spacetime".
This research is supported in part by a Simons Investigator award and National Science Foundation grant PHY-2310429. AL was supported in part by the Stanford Science Fellowship. GT is supported by
CONICET (PIP grant 11220200101008CO),  CNEA, Instituto Balseiro, and a Simons Foundation targeted grant to Instituto Balseiro.

\appendix

\section{Time dependent constrained geometries}
\label{t-dependence}

In this appendix we discuss a generalization of the analysis given in \S
\ref{sec-Constrained-off-shell}, allowing our off-shell configurations to break the $U(1)$ symmetry that was assumed there.  We will find a similar result.

Assuming spherical (or `circular' in the 3d case) symmetry, the most general metric ansatz is 
\be
ds^2=A(\bar{\tau},\bar{r})d\bar{\tau}^2 + B(\bar{\tau},\bar{r}) d\bar{\tau} d\bar{r} + C(\bar{\tau},\bar{r}) d\bar{r}^2 + D(\bar{\tau}, \bar{r}) d\phi^2. 
\ee
Defining a new radial coordinate $r$, we may write $D(\bar{\tau} ,\bar{r})\equiv L^2(r)$ for an arbitrary function $L(r)$.\footnote{The more standard gauge transformation $r^2=D(\bar{\tau} ,\bar{r})$ does not allow non-monotonic variation of the $\phi$-circle size in the radial direction, thus we pick a more general function $L(r)$.} 
By collecting together
terms into new arbitrary functions of $\bar{\tau}$ and $r$ and redefining $A, B, C$ we have
\be
ds^2=A(\bar{\tau},r)d\bar{\tau}^2 + B(\bar{\tau},r) d\bar{\tau} dr + C(\bar{\tau},r) dr^2 +L(r)^2 d\phi^2. 
\ee
We also define a new time coordinate such that 
\be
d\tau=\Phi(\bar{\tau} , r)\left (A(\bar{\tau} , r) d\bar{\tau} +\frac{1}{2} B(\bar{\tau} , r) dr \right )\, ,
\ee
where $\Phi(\bar{\tau} , r)$ is an integrating factor that makes the right-hand side an exact
differential. Squaring $d\tau$ and another redefinition $1/A\Phi^2 \to b^2$ and $C-B^2/4A \to f^2$ yields 
\be
ds^2=b^2(\tau,r) d\tau^2 + f^2(\tau, r) dr^2+ L^2(r) d\phi^2\, .
\ee

The general canonical Euclidean action after imposing the Hamiltonian constraint is (see e.g., \cite{Banihashemi:2022jys} for derivation)
\be
I=\frac{1}{8\pi G_N} \int \! d^3x \, \sqrt{g} \, (K_{ij}K^{ij} -K^2)- \frac{1}{8\pi G_N} \int \! d^2x \, \sqrt{h} \, k-L_0/4G\, ,
\ee
where the `area' term involving $L_0\equiv L(0)$ is present only if there exists a horizon, and the boundary integral is
evaluated at the system boundary. Here $K_{ij}$ is the extrinsic curvature tensor of constant-$\tau$ slices, and $k$ is the trace of the extrinsic curvature of the spatial boundary, as embedded in the constant-$\tau$ surface. 
For the above metric the bulk term vanishes due to spherical symmetry, since $K_{ij} \propto \delta_i^r \delta_j^r$.

The momentum and Hamiltonian constraints are
\be
\frac{L'}{L}\frac{\dot{f}}{f}=0 \quad ; \quad f L''-f' L' + \Lambda f^3 L =0\, .  
\ee
From the first equation we have either $L'=0$ or $\dot{f}=0$; however, a constant non-zero $L$ does not satisfy the second equation. Thus we have $\dot{f}=0$, and by rescaling the radial coordinate we may set $f=1$. So the class of constrained geometries we are considering is
\be
ds^2=b^2(\tau,r) d\tau^2 + dr^2+ L^2(r) d\phi^2/4\pi^2\, ,
\ee
where $L$ satisfies $L''+\Lambda L=0$ and the angular coordinate $\phi$ is rescaled. The above metric differs from \eqref{U1-sym-metric} only by $\tau$-dependence of the $\tau \tau$ component. 
The Euclidean canonical action is still \eqref{I-can-L0}, and has a local maximum at the stationary point $L_0=L_h$. We may assume the function $b$ at the boundary is $\tau$-independent to account for thermal equilibrium, or define the boundary inverse temperature as $\beta_c= \oint d\tau b(\tau,r_c)$. The UV sensitivity and restriction of the integral obtained in \S\ref{sec-UVsensitive}-\ref{subsec:UVcomplete} apply to these more general off-shell configurations.

\bibliographystyle{JHEP}
\bibliography{refs.bib}
\end{document}